\documentclass[a4paper,fleqn]{cas-dc}

\usepackage[numbers]{natbib}

\usepackage{lipsum}
\usepackage{subcaption}

\usepackage{ulem}
\usepackage{xcolor}

\def\tsc#1{\csdef{#1}{\textsc{\lowercase{#1}}\xspace}}
\tsc{WGM}
\tsc{QE}

\begin{document}
\let\WriteBookmarks\relax
\def\floatpagepagefraction{1}
\def\textpagefraction{.001}

\shorttitle{Towards fully automated deep-learning-based brain tumor segmentation: is brain extraction still necessary?}

\shortauthors{Pacheco, Cassia and Silva}  

\title [mode = title]{Towards fully automated deep-learning-based brain tumor segmentation: is brain extraction still necessary?}




%

\author[1]{Bruno Machado Pacheco}[orcid=0000-0002-2635-9340]




\credit{Conceptualization, Software, Validation, Formal analysis, Investigation, Writing - Original Draft, Visualization}

\affiliation[1]{organization={Department of Automation and Systems Engineering},
            addressline={Federal University of Santa Catarina}, 
            city={Florianopolis},
            state={SC},
            country={Brazil}}

\author[2]{Guilherme de Souza e Cassia}[orcid=0000-0001-9029-4679]

\affiliation[2]{organization={Santa Luzia Hospital},
            addressline={D'Or São Luiz Network}, 
            city={Brasilia},
            state={DF},
            country={Brazil}}
            
\credit{Validation, Writing - Review \& Editing}

\author[3]{Danilo Silva}[orcid=0000-0001-6290-7968]


\cormark[1]

\ead{danilo.silva@ufsc.br}


\credit{Conceptualization, Methodology, Resources, Writing - Review \& Editing, Supervision, Project administration}

\affiliation[3]{organization={Department of Electrical and Electronic Engineering},
            addressline={Federal University of Santa Catarina}, 
            city={Florianopolis},
            state={SC},
            country={Brazil}}

\cortext[1]{Corresponding author}



\begin{abstract}[fragile]


State-of-the-art brain tumor segmentation is based on deep learning models applied to multi-modal MRIs.
Currently, these models are trained on images after a preprocessing stage that involves registration, interpolation, brain extraction (BE, also known as skull-stripping) and manual correction by an expert.
However, for clinical practice,  this last step is tedious and time-consuming and, therefore, not always feasible,
resulting in skull-stripping faults that can negatively impact the tumor segmentation quality.
Still, the extent of this impact has never been measured for any of the many different BE methods available.
In this work, we propose an automatic brain tumor segmentation pipeline and evaluate its performance with multiple BE methods.
Our experiments show that the choice of a BE method can compromise up to 15.7\% of the tumor segmentation performance.
Moreover, we propose training and testing tumor segmentation models on non-skull-stripped images, effectively discarding the BE step from the pipeline.
Our results show that this approach leads to a competitive performance at a fraction of the time.
We conclude that, in contrast to the current paradigm, training tumor segmentation models on non-skull-stripped images can be the best option when high performance in clinical practice is desired.

\end{abstract}



\begin{keywords}
Brain Extraction \sep Skull-stripping \sep Brain Tumor Segmentation \sep Deep Learning \sep Evaluation \sep TCGA
\end{keywords}

\maketitle

\section{Introduction}

Precise brain tumor delineation is key to a successful operative plan and post-operative treatment.
Yet, manual delineation of the tumor regions in 3D magnetic resonance images (MRIs) is very laborious \cite{Wang2020}, and the results are subject to intra- and inter-operator variability \cite{Mazzara2004BrainSegmentation,Gordillo2013StateSegmentation}.
Therefore, there has been significant interest in automatic brain tumor segmentation, for which deep learning models have shown great success. \cite{Menze2015TheBRATS}

\begin{figure*}[t]
    \centering
    \includegraphics[width=\textwidth]{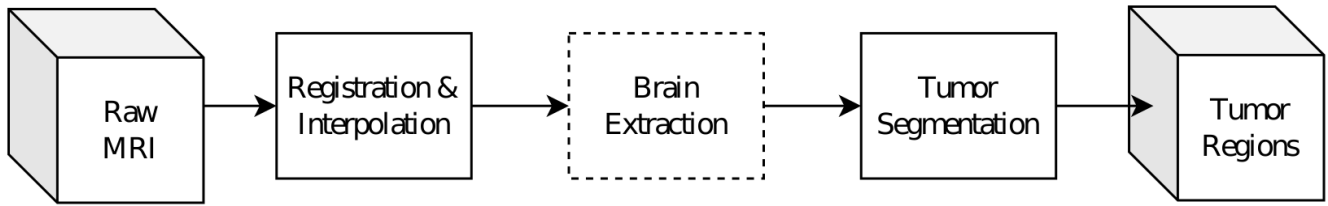}
    \caption{Pipeline of operations for tumor segmentation on raw brain MRIs. The Brain Extraction step is highlighted as it may require manual correction.}
    \label{fig:brats-pipeline}
\end{figure*}

Due to the many MRI settings available, two images of the same subject can show the same tissue with drastically different appearance, which imposes a great challenge for deep learning models.
The current paradigm is to preprocess the brain MRIs before feeding them to the tumor segmentation model, performing: registration (to a template), interpolation, skull-stripping---also known as brain extraction (BE)--- and manual correction of the brain mask \cite{Bakas2018IdentifyingChallenge}.
The first two operations aim to achieve spatial normalization for images acquired in different machines, with different resolutions, etc.
BE removes unnecessary information for the tumor segmentation, namely the skull and other non-brain tissues that also appear in the MRIs.
The last step, manual correction, is the intervention of experts to correct defects from the BE step, once BE methods occasionally leave non-brain voxels in the images or remove brain voxels from them.
The pipeline of operations required to generate the tumor segmentation from raw MRIs is illustrated in figure \ref{fig:brats-pipeline}.

In a fully automated solution, which is desirable in clinical practice, the manual correction during the preprocessing has to be avoided.
In this case, the images fed to the tumor segmentation model still contain the possible defects from the BE methods.
As brain tumor segmentation models are trained on gold-standard images, these defects are unexpected and, thus, may hinder the quality of the resulting segmentation.
This is why it is crucial to the brain tumor segmentation pipeline that the right BE method is selected from the many options \cite{Smith2004AdvancesFSL,Segonne2004AMRI,Kamnitsas2017EfficientSegmentation,Isensee2019AutomatedNetworks,Thakur2020BrainTraining}.

\begin{figure*}[T]
    \centering
    \begin{subfigure}[b]{0.45\textwidth}
        \centering
        \includegraphics[width=\textwidth]{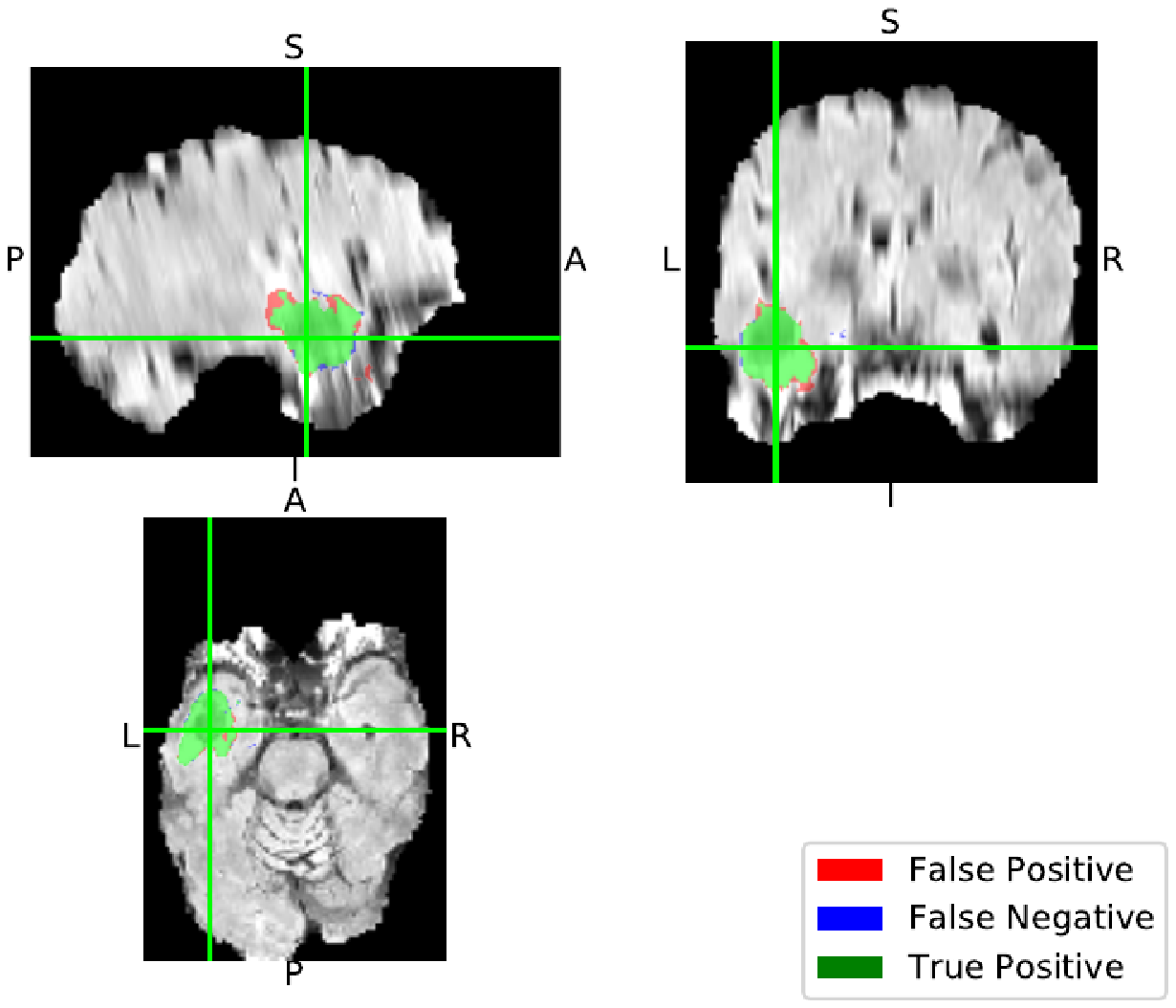}
    	\caption{}\label{fig:bet-impact-example-a}
    \end{subfigure}
    \hfill
    \begin{subfigure}[b]{0.45\textwidth}
        \centering
        \includegraphics[width=\textwidth]{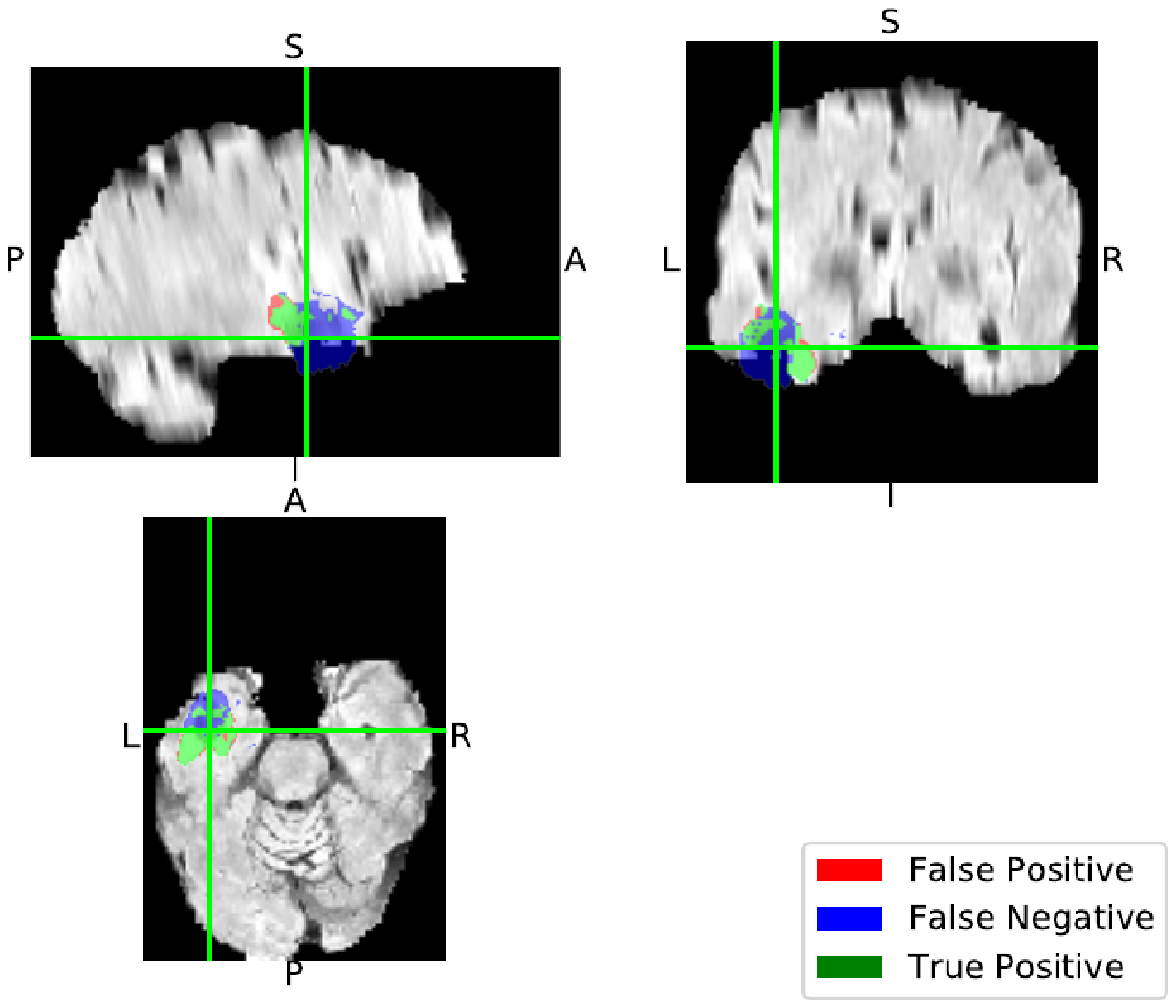}
	\caption{}\label{fig:bet-impact-example-b}
    \end{subfigure}
    \caption{
    T2-Flair 3D MRI (orthogonal view) overlaid with predicted Whole Tumor (WT) segmentation. In a) the image was skull-tripped with BrainMaGe\cite{Thakur2020BrainTraining} (BE Dice score of 0.92), which missed few brain voxels but left many non-brain voxels. 
    In b), the image was skull-stripped with DeepMedic\cite{Kamnitsas2017EfficientSegmentation} (BE Dice score of 0.96), missing few brain voxels from the tumor region. The tumor segmentation model\protect\footnotemark {} produced a much better segmentation 
    on a) than on b) (Dice scores of 0.80 and 0.57, resp.), as one can see in the coloring, missing a big portion of the tumor on the image skull-stripped with DeepMedic.}
    \label{fig:bet-impact-example}
\end{figure*}

Yet, to the best of our knowledge, no one has quantified the impact of BE methods on state-of-the-art tumor segmentation deep learning models.
In practice, the performance on the BE task (how well a method distinguishes brain voxels from non-brain voxels in the image) is used as a guideline to choose one method over the other.
However, BE performance does not translate directly to low-impact in tumor segmentation.
For instance, a BE method that misses very few brain voxels but makes these mistakes in the tumorous region will have a higher impact in the tumor segmentation than a BE method that misses more brain voxels but far from the tumor.
This is because the images resulting from the first method will miss tumor context, which is essential for tumor segmentation models.
Fig. \ref{fig:bet-impact-example} shows a real example of how two high-performant BE methods can have significantly different impacts in a tumor segmentation model.

In this work, we provide the first quantitative evaluation of BE methods as a preprocessing step for tumor segmentation.
We compare the most commonly used BE methods with respect to their impact in the downstream task.
To enable this evaluation, we develop and validate a preprocessing pipeline that aligns raw images to available tumor annotations.
This is necessary because publicly-available datasets of brain MRIs provide tumor annotations only for the preprocessed images, whereas our evaluation must be performed on raw images.
Our results show that up to 15.7\% of the tumor segmentation performance can be compromised by a poorly-selected BE method.
Furthremore, the BE methods that have low impact in the tumor segmentation performance present longer inference time and impose higher computational costs.
In face of this, we investigate whether BE is necessary at all for tumor segmentation.
We note that the best performing BE methods are all based on deep learning models.
Therefore, we hypothesize that a single deep learning model can incorporate both tasks, an approach shown to be successful for BE and brain tissue segmentation \cite{Cullen2018}.
We achieve this by training tumor segmentation models using only non-skull-stripped images, which is possible due to the preprocessing pipeline that we developed for the BE evaluation.
Our experiments with this single-stage approach indicate that it can achieve competitive performance while being one of the fastest solutions.
This points towards a new paradigm for developing brain tumor segmentation models when suitability for clinical practice is desired.

\footnotetext{The tumor segmentation model used was a 3D U-Net as described in sec. \ref{sec:tumor-segmentation}.}

In summary, our contributions are:
\begin{itemize}
    \item a brain tumor segmentation pipeline that requires no manual intervention to generate high-quality tumor segmentation (sec. \ref{sec:methods});
    \item an evaluation of the impact of BE methods on tumor segmentation performance and computational cost (sec. \ref{sec:tumor-segmentation});
    \item a single-stage deep learning model for tumor segmentation of non-skull-stripped images (sec. \ref{sec:brats-no-bet}) 
\end{itemize}


\section{Related Work}

Multiple evaluations of BE methods on MRIs have been reported.
\citet{Isensee2019AutomatedNetworks} proposed a deep learning model and compared it with 6 publicly available choices in multiple neuroimaging datasets.
\citet{Thakur2019Skull-StrippingLearning} presented an evaluation of deep learning BE methods (including the one they propose) and a thorough evaluation of the usefulness of the different MRI modalities.
These two works have significant relevance for the context of tumor segmentation, as they presented performant models on images of pathologically-affected brains.

At the same time, works in fully-automatic brain tumor segmentation are scarce and datasets that support these experiments are few.
\citet{Kamnitsas2017EfficientSegmentation} reported results on a private dataset that underwent automatic preprocessing, but did not discuss the role of the BE method used, nor reported results with different BE methods.
\citet{Uhlich2018} evaluated different BE methods but did so on the segmentation of fewer tumor regions (whole tumor and edema) and with only T1-weighted and FLAIR modalities, which is not directly comparable to the state-of-the-art results of the BraTS competition \cite{Menze2015TheBRATS,Bakas2018IdentifyingChallenge}
\citet{yogananda2019} performed automatic preprocessing on brain MRIs, but for the task of brain segmentation (segmenting brain images into gray matter, white matter and cerebrospinal fluid).
They have also trained the models exclusively for BE, in a similar approach to using a BE method publicly available.
In a concurrent work to ours, \citet{kondrateva2022do} have presented a short ablation study on the effects of preprocessing operations on brain tumor segmentation, including skull-stripping, and achieved similar conclusions to ours with respect to the utility of BE as a preprocessing step for tumor segmentation.
Yet, the authors trained their models on fewer images and lacked data augmentation on their experiments, besides using a single BE method.
In fact, they have not disclosed which BE method they used in their experiments.

To the best of our knowledge, no work has been presented in the literature that focuses on the BE impacts for state-of-the-art tumor segmentation.
Instead of proposing another BE method, this work is the first to evaluate publicly available BE methods and present their application in a fully-automatic setup, suitable for clinical practice.
Furthermore, we also compare the automatic BE methods with the approach of not performing BE at all, training tumor segmentation models on images without skull-stripping.

\section{Background}

\subsection{Brain Tumor Segmentation}

Even though brain (and other nervous system) tumors account for a small share of total tumor incidences, they have a disproportional death rate and are the leading cause of cancer death among young (<20 years) males and females \cite{Miller2021Brain2021}.
Among such malignancies, gliomas have the highest incidence rate \cite{Holland2001ProgenitorFormation}.
Specially for these types of tumors, magnetic resonance imaging is a powerful tool and is one of the most common tests for their diagnosis, survival prediction, and surgery planning \cite{Havaei2017BrainNetworks,Myronenko20183DRegularization}.

In these images, the segmentation of tumor regions can improve diagnosis, growth rate prediction, treatment planning, and treatment response assessment \cite{Havaei2017BrainNetworks}.
Usually, proper tumor segmentation involves the distinction between healthy tissue and the glioma sub-regions, i.e., peritumoral edema/invaded tissue, necrotic and other non-enhancing regions, and enhancing core \cite{Bakas2017AdvancingFeatures}.
Unfortunately, segmenting such tumors is a strenuous task, as gliomas come in varied shapes and sizes, appear in multiple regions, and are often poorly contrasted.
Furthermore, as infiltrative tumors, they are poorly marginated and have ill-defined borders.
For these reasons, gold-standard segmentation is typically done with multimodal MRI.

\subsubsection{Automatic Segmentation}

MRI scans can be acquired in multiple machines with different acquisition parameters and protocols.
This results in images from different hospitals showing the same tumorous cells with drastically different appearances.
Tumor segmentation algorithms must be robust to this variability, as they are expected to perform on multi-institutinal datasets.
Therefore, a very intricate preprocessing pipeline would be necessary for traditional computer vision algorithms, whereas learning-based approaches thrive, given enough data.

Even though many algorithms were developed to perform fully-automatic brain tumor segmentation \cite{Gordillo2013StateSegmentation}, this task has seen great improvements since deep learning models were first trained for it \cite{Bakas2018IdentifyingChallenge}.
In fact, deep learning models have become state-of-the-art for many cancerous tissue segmentation tasks, even on different imaging tests \cite{Huang2021Rectal,Fan2020LiTS,SINGH2020112855,Woniak2021,Dash2022}.
Convolutional Neural Networks (CNNs) are naturally useful for the task as they excel at capturing spatial information with fewer parameters than fully connected networks.
CNNs have been the choice for many of the state-of-the-art models over recent years.
\citet{Myronenko20183DRegularization} has proposed to train a CNN with a variational auto-encoder branch to reconstruct the input image as a second output, thus, regularizing the features of the network's encoder.
\citet{McKinley2018EnsemblesSegmentation} combined the predictions of multiple dense CNNs, which were trained using a modified loss function accounting for uncertainty and noise in the ground truth labels.
\citet{Zhao2019BagSegmentation} applied modifications common to many different applications area to the brain tumor segmentation task, attacking the imbalance of the labels, extending the training data through semi-supervised learning, combining models and predictions, and optimizing the training method.

The U-Net and its 3D version \cite{Ronneberger2015U-Net:Segmentation,Cicek20163DAnnotation} are widely used CNN architectures designed for medical imaging and have shown great results in this domain, stablishing a design pattern that is followed by state-of-the-art models such as the ones referenced above.
\citet{Isensee2020NnU-Net:Segmentation} propose the use of expert knowledge to properly train a U-Net in spite of complex architectural changes, naming this network nnU-Net \cite{Isensee2018NoNew-Net}.
Upon dataset characteristics, its heuristics define pre-processing operations, network architecture, hyperparameters, and training approach.
Ever since its remarkable performance in the 2020 edition of the brain tumor segmentation challenge \cite{Isensee2020NnU-NetSegmentation}, nnU-Net has become a reference for brain tumor segmentation.

\subsubsection{BraTS Challenge}\label{sec:brats-challenge}

The brain tumor segmentation challenge (BraTS) aims to benchmark state-of-the-art tumor segmentation algorithms in multi-modal, pre-operative brain MRIs \cite{Menze2015TheBRATS,Bakas2018IdentifyingChallenge}.
Organized yearly, the competition gathers images from several institutions and assembles the largest publicly available dataset of brain scans with complete tumor annotation.
Through this public dataset, competitors train and validate their models, but are ranked on their performance on a non-disclosed test set.
This way, BraTS has become the main reference for brain tumor segmentation performance, providing a basis of comparison for all improvements in the field.

The data provided since 2017 contains pre-operative scans of patients with confirmed glioma (low grade or high grade).
For each subject, the multimodal MRI scans describe T1-weighted, contrast-enhanced T1-weighted, T2-weighted, and T2-Flair modalities, all of them already preprocessed, with manual tumor segmentation by experts.

The tumor segmentation masks consist of annotations of the enhancing tumor, the peritumoral edema, and the necrotic and non-enhancing regions of the tumor.
Still, following the competition’s evaluation criteria, it is very common (and we follow this procedure) to segment and to analyze: the whole tumor (WT), which comprises all the annotated regions; the tumor core (TC), which is the union of the enhancing tumor region and the necrotic and non-enhancing regions; and the active tumor (AT), which comprises the enhancing tumor regions as segmented.

To evaluate each competitor, BraTS resorts to computing two measures (Dice coefficient and Hausdorff distance, see section \ref{sec:seg-metrics}) for each segmented region (WT, TC, and AT) of each subject in the test set.
Then, the competitors are ranked in each combination of metric and region, resulting in 6 individual rankings for each patient.
Finally, the rankings for each subject are averaged, and the resulting values across all subjects are averaged for each participant, resulting in a final ranking score, which is used to compute the final ranking of the competitors. \cite{Menze2015TheBRATS,Bakas2018IdentifyingChallenge}

\subsection{Brain MRI Preprocessing}\label{sec:preprocessing}

Due to the many sequences that exist to acquire MRI, it is very unlikely that an algorithm developed on images from a given institution will work properly on images from a different source.
Therefore, it is common practice (in semantic segmentation tasks on brain imaging) to perform some preprocessing operations that aim to standardize images with regard to their shapes, spatial orientation, resolution, intensities, and other parameters \cite{Amorosino2022}.
For brain tumor segmentation, the preprocessing operations most often performed are registration, interpolation, and brain extraction \cite{Bakas2018IdentifyingChallenge,Akkus2017}.

Registration means to find a transformation that, when applied to a given image, aligns it to another \cite{Oliveira2012}.
This transformation can be found by optimization, that is, by minimizing some quantitative measure of the misalignment between the two images.
In the context of brain MRIs, registration is applied both to align the multiple modalities acquired from the same patient and to align images from different patients to a common reference.
Oftentimes performed with the registration, interpolation changes the resolution of the images so they are not only spatially-aligned but also aligned voxel-wise.

BE is one of the most applied preprocessing steps for the analysis of brain MRI \cite{bloomfield2017basic}.
Also known as skull-stripping, it (ideally) removes all non-brain tissues from the image.
BE is usually performed by replacing all voxels that contain such tissues with background, effectively leaving only the brain information in the image.
For this reason, it can be seen as a segmentation task for which the brain mask is the target.

Finally, many other preprocessing operations can be performed in brain MRI beyond the 3 already described.
Normalization, which homogenizes the distribution of voxel intensity across different patients, is usually performed when it is not part of the following steps in the pipeline.
Bias field correction is also a preprocessing step to fix this commonly present artifact in brain MRI, for which the N4 algorithm \cite{Tustison2010N4ITK:Correction} is a standard tool.

\subsubsection{Brain Extraction Methods}\label{sec:brain-x}

The BE task presents several challenges, as brain shapes and sizes vary greatly from patient to patient.
For this reason, manual segmentation is still considered the state-of-the-art in BE.
Still, it is very costly as it takes plenty of time from a senior neuroradiologist to fully annotate a brain in a 3D MRI with high accuracy.  
Yet, even experts are challenged by the segmentation of certain parts of the cranial sinuses and the cerebellum \cite{KLEESIEK2016460}.

Several automated BE methods have been proposed through the years.
First with well-defined, classical computer vision algorithms \cite{Smith2002FastExtraction,Segonne2004AMRI}, and, more recently, with deep learning approaches \cite{Bakas2018IdentifyingChallenge}.
Even though much more consistent than the human experts, the automated methods suffer greatly from differences in the acquisition parameters of the MRI scans and from the presence of brain lesions.
Even though these shortcomings have been the subjects of recent research, no BE method has been elected as the go-to approach, as computational cost and overall performance across different applications are still hurdles.
Below, we review the most popular fully-automatic methods.

\paragraph{BET}

The Brain Extraction Tool (BET), available as part of FSL (FMRIB Software Library) \cite{Smith2004AdvancesFSL}, is one of the most cited BE tools.
Based on an evolutionary deformable model, this is a lightweight approach developed to be used on T1-weighted images.
Popular variations of BET are also packaged in FSL.
The variation that presents better results for the images of interest in this study is one that runs also multiple bias field removal operations to improve the resulting brain mask.
We will refer to this variation as "reduce bias".

\paragraph{FreeSurfer}

A mix of a deformable model and a watershed algorithm \cite{Segonne2004AMRI}, FreeSurfer’s skull-stripping method tops with BET as one of the most commonly used tools for the task.
It was also developed to work on T1-weighted MRIs.

\paragraph{DeepMedic}

DeepMedic \cite{Kamnitsas2017EfficientSegmentation} was designed for brain lesion segmentation and won the ISchemic LEsion Segmentation (ISLES) challenge \cite{Winzeck2018ISLESMRI}.
As one of the first 3D architectures to emerge in the area, it contained an 11-layers deep, double pathway 3D CNN.
Later, DeepMedic was trained for BE and made available through the Cancer Imaging Phenomics Toolkit (CaPTk) \cite{Davatzikos2018CancerOutcome,Pati2020TheOverview}.
In the toolkit, two models are available: a modality-agnostic model, which was trained on the four modalities without distinction, and a multi-modality model (henceforth called "multi-4", following the authors’ naming), which has one input channel for each of the four modalities.

\paragraph{HD-BET}

HD-BET's approach to skull-stripping builds upon a 3D U-Net with pre-activation residual blocks in the encoder \cite{Isensee2019AutomatedNetworks}.
The network was trained on a multi-institutional dataset (EORTC-26101) containing the same 4 modalities as in the BraTS dataset.
The training was performed in a modality-agnostic approach, using the four modalities available.
With this, HD-BET reported the best results thus far in publicly available datasets.

HD-BET defaults to using an ensemble of five models and performing test-time augmentation (mirroring), which results in high inference times.
It offers, though, a "fast" option, which uses only one of the models and does not perform test-time augmentation.

\paragraph{BrainMaGe}

The development focus of BrainMaGe was in brains with apparent pathologies \cite{Thakur2020BrainTraining}.
3D-Res-U-Nets were trained in a multi-institutional private dataset in a modality-agnostic approach and a multi-modality approach (using the modalities as channels of the input), similar to DeepMedic.

\subsection{Segmentation metrics}\label{sec:seg-metrics}

The most commonly used metrics to evaluate the quality of a semantic segmentation (e.g., BE, tumor segmentation) are the Dice coefficient and the Hausdorff surface distance. The first, also known as Sørensen-Dice, compares the intersection of the predicted segmentation $P$ and the ground-truth $T$ to the sum of positive voxels in each:
\begin{equation*}
\mathrm{\textit{Dice}} = \frac{2\left|P \cap T \right|}{\left| P \right| + \left| T \right| }.
\end{equation*}
Therefore, it is always limited to the $[0,1]$ interval (the higher the better).

The Hausdorff distance, on the other hand, measures the smallest distance between any point in a given surface and all the points in another surface, returning the biggest value found for all the points.
It measures the surface distance between the segmentations (ground-truth vs predicted), that is, it calculates the biggest gap there is between both surfaces
\begin{equation*}
    \mathrm{\textit{Hausdorff}}\left( P,T \right) = \max\left\{ \sup_{p \in \partial P} \inf_{t \in \partial T} d( p,t ) , \sup_{t \in \partial T} \inf_{p \in \partial P} d( t,p ) \right\}
\end{equation*}
where $d\left( \cdot,\cdot \right) $ is the euclidean distance, and $\partial P$ and $\partial T$ are the surfaces of the segmentations.
To ensure stability, the 95th percentile of the distances is used instead of the maximum.
This approach will be henceforth referred to as \textit{Hausdorff95} (HD95).
HD95 is always a non-negative value and only returns 0 when the segmentations are identical.

\section{Methods}\label{sec:methods}

\subsection{Data}\label{sec:data}

\begin{figure*}[T] 
    \centering
    \includegraphics[width=0.7\textwidth]{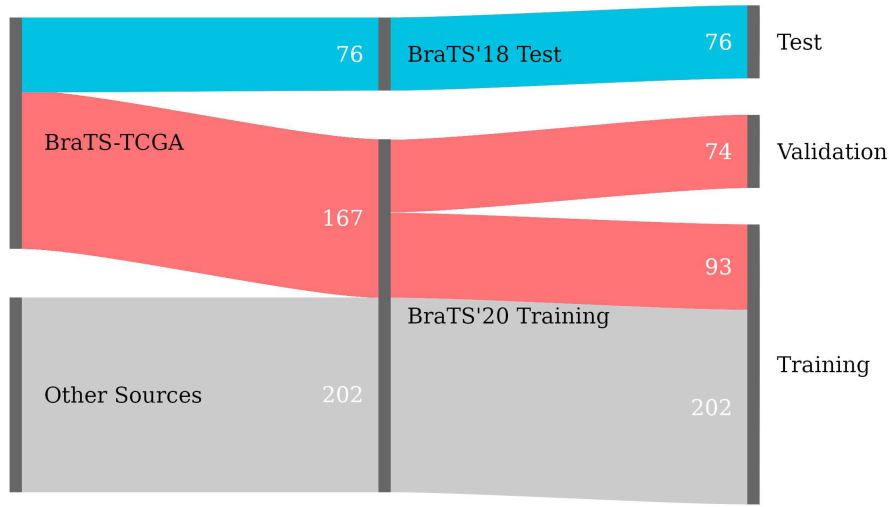}
    \caption{Composition of the datasets used in the experiments. Gray links represent images that are only available after preprocessing and skull-stripping (BraTS-preprocessed), while colored links represent images that are available both raw and as preprocessed by BraTS. 
    For images represented by red links, the transformation that aligns the raw version to the BraTS-preprocessed version is also available
    (through DICOM-Glioma-Seg).}
    \label{fig:datasets}
\end{figure*}

In order to evaluate the performance of a BE method on the downstream task of tumor segmentation, raw images (i.e., before skull-stripping) are needed together with their corresponding tumor segmentation.
However, the data provided by BraTS include only skull-stripped images.
Fortunately, two of the sources of the BraTS data are the TCGA-LGG and TCGA-GBM collections, which provide the raw images together with subject IDs that are retained in the BraTS data.
For convenience, we refer to the subset of the TCGA-LGG and -GBM collections that is used in the BraTS challenge as the BraTS-TCGA data \cite{Bakas2017SegmentationSet,Bakas2017SegmentationSetb}.

The BraTS-TCGA data contains, for each of the 243 subjects, three images: raw MRI, BraTS-preprocessed\footnotemark MRI, and tumor segmentation.
\footnotetext{"BraTS-preprocessed MRIs" refers to the images that were preprocessed by the competition organizers, following their preprocessing pipeline.}
Of the 243 subjects, 76 were part of the test set of the 2018 edition of BraTS (BraTS'18), while the remaining 167 subjects were on the training set of the 2020 edition of BraTS (BraTS'20).
The training set from BraTS'20 also contains 202 subjects from other sources (for a total of 369 subjects), whose corresponding raw images are not available.

We split the available images in 3 sets: Training, Validation and Test.
This way, we can tune hyperparameters on models trained on the Training set and evaluated on the Validation set, while keeping the Test set untouched.
Then, we can retrain the models using the tuned hyperparameters on the combination of Training and Validation sets, and perform a final evaluation on the Test set.
The construction of our datasets is illustrated in figure \ref{fig:datasets}.

First and foremost, we build the Test set from the raw images and tumor segmentation labels of the 76 subjects that were part of the test set of BraTS'18.
This guarantees that the Test set does not contain any image used in the development of the BE methods.
From the remaining subjects of BraTS-TCGA, we randomly choose 74 to form the Validation set, again taking the raw images and corresponding tumor segmentation labels.
Finally, the Training set consists of the BraTS-preprocessed images and tumor segmentation labels for the 295 subjects that are not part of the Test and Validation sets.

This way, we build a training set in the same manner as the one used for the participants for the competition.
With such a training set, we can replicate the training of state-of-the-art models for tumor segmentation.
But most importantly, we build Validation and Test sets with raw images, which are necessary to simulate the requirements of clinical practice.
This allows us to verify the correctness of our implementation, as well as to guide our decisions on which BE methods and configurations to adopt and compare.

\subsection{Preprocessing}

To ensure a proper evaluation, we follow the preprocessing of the BraTS data for the raw images.
The operations applied by the BraTS competition to the MRIs are: registration to SRI-24 brain atlas, interpolation to an isometric 1 mm³ resolution, and skull-stripping \cite{Bakas2018IdentifyingChallenge}.
For all TCGA-LGG and TCGA-GBM that were used in the BraTS'20 training set, the transformations that align the raw images to the BraTs-preprocessed images (and, thus, the atlas) are available in the DICOM-Glioma-SEG dataset \cite{Beers2018}.
Therefore, we use these transformations instead of performing the registration operation.
We used SimpleITK \cite{Yaniv2018} to apply the transformations and to interpolate the results.

\begin{figure*}[t]
    \centering
    \includegraphics[width=\textwidth]{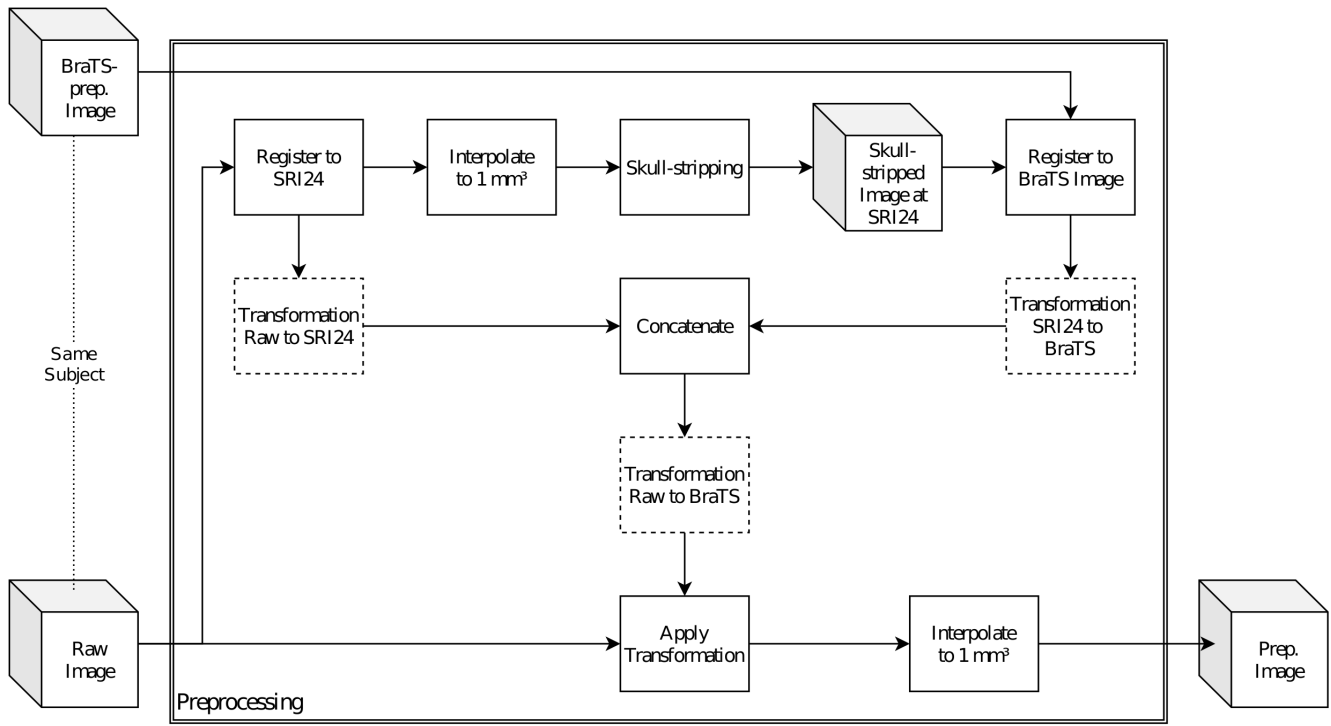}
    \caption{Preprocessing pipeline diagram. The resulting image is not only registered and interpolated in the same way as the BraTS data, but also aligned to it. The only difference is that the resulting image is not skull-stripped.}
    \label{fig:preprocessing-pipeline}
\end{figure*}

Yet, for the images in Test set (BraTS'18 Test) these transformations are not available.
Thus, a more complex preprocessing pipeline is necessary.
Following the recommendation of the competition’s organizers, we replicate this pipeline using CaPTk \cite{Davatzikos2018CancerOutcome,Pati2020TheOverview}, which uses Greedy \cite{Battle2016IC-P-174:MRI} for registration.
Our experiments with different registration tools resulted in poorer tumor segmentation results in the Validation set.
We do not normalize the images, as that is already performed by nnU-Net.
Bias field correction using N4 showed no overall improvements in the Validation set performance of BE and tumor segmentation, therefore we do not use it.
The preprocessing pipeline is pictured in figure \ref{fig:preprocessing-pipeline}.

Making sure that the Test set images preprocessed by us are aligned to the BraTS-preprocessed images is crucial as it allows us to use the tumor segmentation labels without any modifications or transformations and, therefore, no quality loss.
For this, after the registration to SRI24 and interpolation, we align the images to their corresponding BraTS-preprocessed image.

Registering a non-skull-stripped image to a brain-only image could result in non-ideal transformations.
Therefore, we also perform skull-stripping on the images after registration to SRI24 and interpolation, and before registration to the BraTS-preprocessed image, as shown in figure \ref{fig:preprocessing-pipeline}.
We use HD-BET for this operation.
Note that the BE performed with HD-BET is used only to improve the SRI24-to-BraTS transformation, thus, it leaves no traces in the final preprocessed image (which is not skull-stripped), therefore, no bias is inserted in favor or against the BE method during the preprocessing.

Note that applying the transformation generated by the second registration operation to the already registered image would require yet another interpolation, which would lower the image quality.
Therefore, we concatenate both transformations generated, resulting in a single rigid transformation that aligns the original raw image to the BraTS-preprocessed image.
This final transformation is followed by an interpolation to the target resolution (1 mm³).

Our preprocessing pipeline results in images aligned to the BraTS-preprocessed images of the same subject, which are aligned to SRI-24, and interpolated to 1 mm³ resolution.
The only difference between our images and the BraTS-preprocessed ones is that our images still contain skull and other non-brain tissues, i.e., not skull-stripped.
This means that the application of a BE method in such images results in an image very similar to the corresponding BraTS-preprocessed image (even aligned to the ground-truth tumor segmentation), but with a different brain mask.

In summary, we preprocessed all BraTS-TCGA images that were part of the BraTS'20 Training set (see figure \ref{fig:datasets}) using the transformations available in the DICOM-Glioma-Seg dataset.
This already guarantees that the resulting images are aligned to the BraTS-preprocessed images.
The remaining BraTS-TCGA images, which compose the Test set, were preprocessed with the pipeline explained above and illustrated in fig. \ref{fig:preprocessing-pipeline}.

\subsection{Brain Extraction}

From the many BE tools available, we selected some popular choices in works on brain tumor segmentation, by number of citations.
These are the tools presented in section \ref{sec:brain-x}.
Even though some methods are modality-agnostic, we evaluate them in the T1-weighted modality, which is the modality for which BET and FreeSurfer were designed.
Besides that, we also evaluate DeepMedic's and BrainMaGe's multi-4 models, which are applied to all four MRI modalities at once.

We achieved better results on the Validation set by applying BE after both registration and interpolation, as the input image has a better resolution and we do not need to interpolate the brain mask.

\subsection{Brain Tumor Segmentation}\label{sec:brats-models}

\begin{figure*}[T]
    \centering
    \begin{subfigure}[b]{.9\textwidth}
        \centering
        \includegraphics[width=\textwidth]{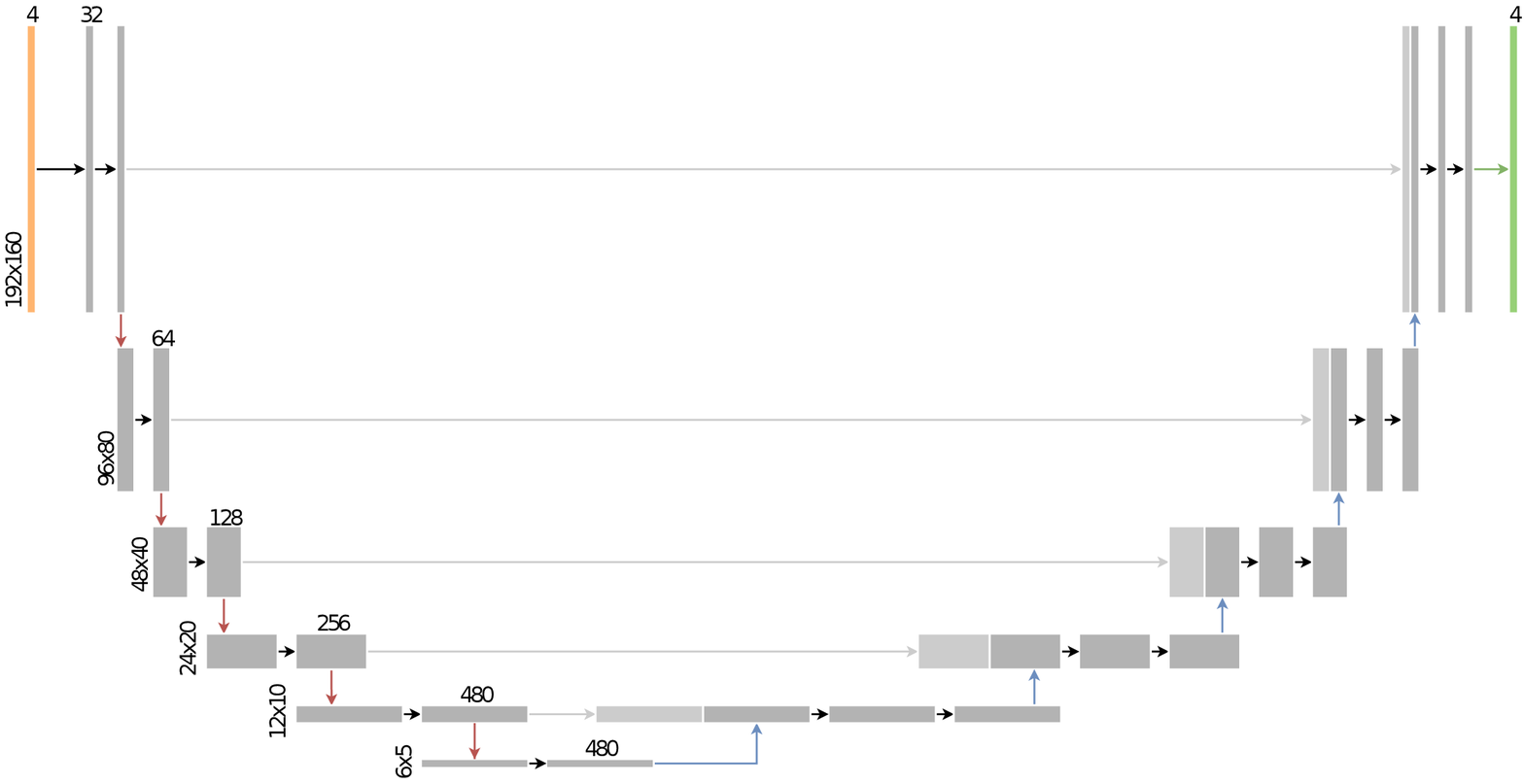}
    	\caption{}\label{fig:unet-2d}
    \end{subfigure}
    \begin{subfigure}[b]{.9\textwidth}
        \centering
        \includegraphics[width=\textwidth]{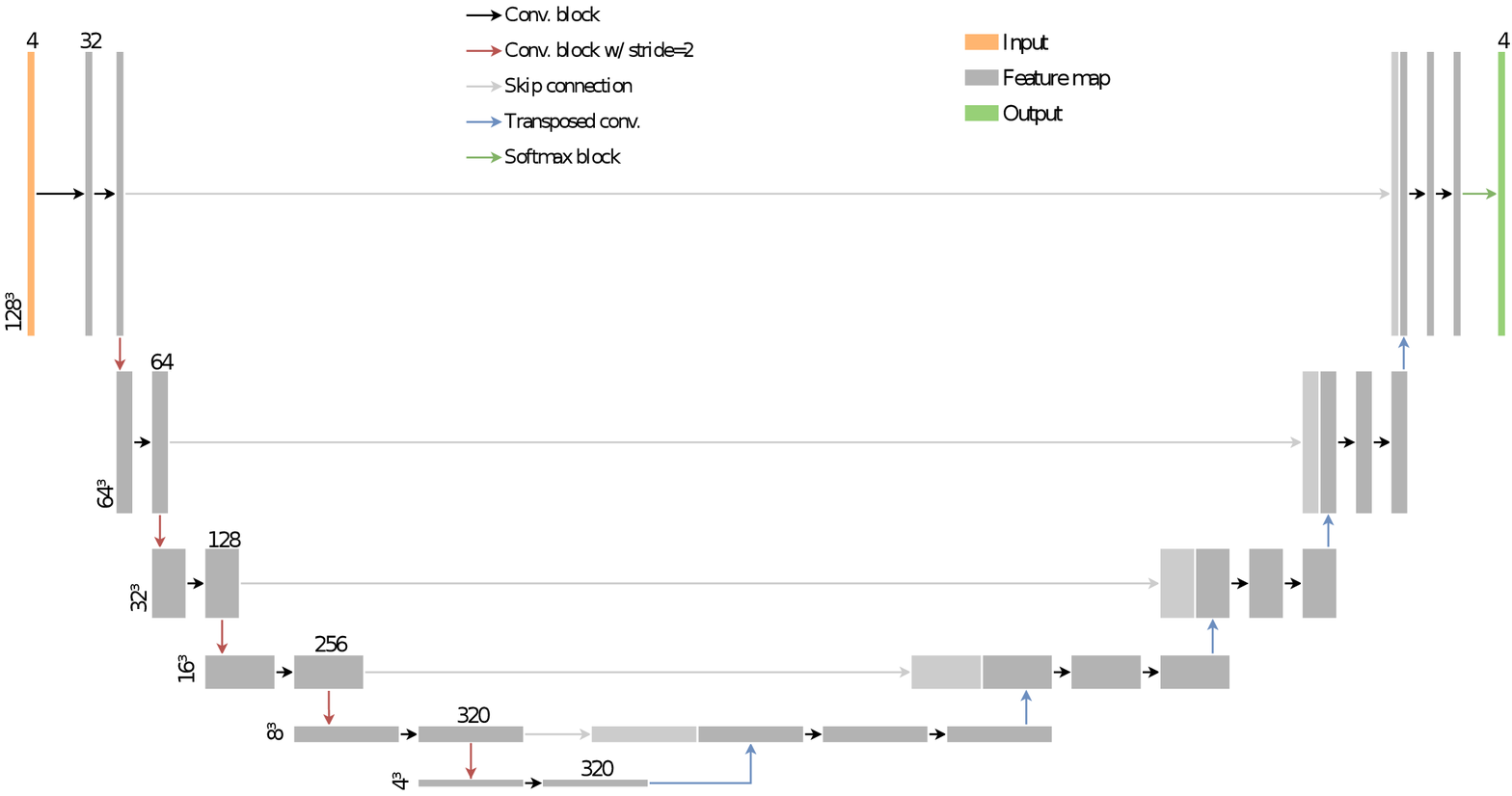}
    	\caption{}\label{fig:unet-3d}
    \end{subfigure}
    \caption{Architectures of the 2D (\ref{fig:unet-2d}) and 3D (\ref{fig:unet-3d}) U-Nets trained with nnU-Net, as described in sections \ref{sec:brats-models} and \ref{sec:tumor-segmentation}. On the left of the feature map blocks are the dimensions, and in the top are the number of channels. These two parameters are uniform horizontally (i.e., these are the same for all blocks side-by-side). Convolution blocks are composed of a convolution operator (kernel of size 3 in all dimensions, unitary stride and padding), followed by batch normalization and LeakyReLU. Transposed convolutions have kernel of size 2 (in all dimensions), stride of 2 (in all dimensions) and unitary padding. Softmax block is composed of a convolution operator as in the convolution block followed by softmax across image channels.}
    \label{fig:unet-architectures}
\end{figure*}

We use the nnU-Net as our reference for tumor segmentation.
Given its excellent performance in the past editions of BraTS, it has proven to be a solid deep-learning approach to tumor segmentation.
Using nnU-Net, we traing both 2D and 3D U-Nets.
Details of the architecture can be seen in figure \ref{fig:unet-architectures}.
Even though the 3D model has shown better results in the competition, the 2D model is much lighter (requires less computational resources) to train and to perform inference.

We train the U-Net models mostly following the implementation of the nnU-Net authors, which was developed based on the BraTS'20 Training data \cite{Isensee2020NnU-NetSegmentation}.
We ran region-based training, which means that the network is trained on the regions used by the competition to evaluate the models (WT, TC, and AT), instead of the original regions provided in the segmentations (enhancing tumor, peritumoral edematous/invaded tissue, and necrotic tumor core).
Batch normalization was used as suggested by the authors of BraTS.
We use the data augmentation of nnU-Net \cite{Isensee2018NoNew-Net}, namely: rotation, scaling, elastic deformation, gamma correction and mirroring.


\subsection{Evaluation Metrics}\label{sec:evaluation}

We stick to the two main metrics used by BraTS: the Dice coefficient and the Hausdorff surface distance \cite{Menze2015TheBRATS}.
With 2 metrics and 3 segmentation regions, the comparison of models becomes troublesome.
BraTS solves this problem through a ranking scheme.
However, we aim to evaluate the performance of the approaches rather than to pick the most successful one.
Therefore, we opt to summarize each metric instead of ranking them.
More specifically, we first average the score on the 3 regions for each image and for each metric.
Then, we take the median value over the images, resulting in two scores: a median average Dice and a median average Hausdorff95.

We also compute the inference time of each BE method and each nnU-Net model.
We consider the inference time as the wall time each method took from reading the input image(s) to generating (and storing) the segmentation (tumor segmentation or brain mask).

\section{Experiments and Results}

To evaluate the BE methods presented in the previous section in the task of tumor segmentation and experiment brain tumor segmentation without any BE method, we present in this section 3 sets of experiments.
First, in \ref{sec:preprocessing-experiments}, we test whether our preprocessing pipeline is successful in aligning the raw brain MRIs to the tumor annotations provided for the BraTS-preprocessed images, as discussed in \ref{sec:data} and \ref{sec:preprocessing}.
Then, in \ref{sec:tumor-segmentation}, we train brain tumor segmentation models and apply them to images preprocessed with the different BE methods.
Finally, in \ref{sec:brats-no-bet}, we train and test tumor segmentation models on images preprocessed without skull-stripping.

All experiments reported below were performed on a Linux machine with 16 CPUs, 32 GB of memory, an RTX 2080 Ti (11 GB), and a solid-state drive.
We did not restrict the hardware usage of any method or model, which means that most of them used the GPU and as many threads as possible.
In fact, the only methods that used solely the CPU were BET and FreeSurfer.

\subsection{Preprocessing}\label{sec:preprocessing-experiments}

To validate that the images resulting from our preprocessing pipeline (see sec. \ref{sec:preprocessing}) are aligned to the BraTS-preprocessed images, we quantitatively evaluated the similarity between the two.
More specifically, for each modality of each image, we computed Pearson's correlation and peak signal-to-noise ratio (PSNR) between the ones preprocessed by us and the BraTS-preprocessed ones.
Since our images contain non-brain voxels as well, we limit our analysis to the regions that are not background in the BraTS-preprocessed images (i.e., only the brain region), applying this mask to our images before the computation.

On the Test set, the average correlation was 0.94 and the average PSNR was 37\,dB.
For comparison, performing this evaluation on BraTS-preprocessed images after a random 1\,mm translation in any orthogonal direction results in an average correlation of 0.89 and an average PSNR of 27\,dB, confirming the similarity between the images preprocessed by us and the BraTS-preprocessed images.


\subsection{Brain extraction for tumor segmentation}\label{sec:tumor-segmentation}

We trained the tumor segmentation models (see sec. \ref{sec:brats-models}) on the BraTS-preprocessed Training set images and evaluated them on the Validation set images after our preprocessing.
We first trained the models for 1000 epochs on the Training set and defined the total number of epochs based on the Validation set performance (see sec. \ref{sec:data} for the definition of the sets); specifically, the total number of epochs is chosen as the smallest value such that the model achieves its peak validation performance.
The 2D model converged within 150 epochs to its peak performance on the Validation set, while the 3D model required 500 epochs.

The final models are then trained on the union of Training and Validation sets for the number of epochs defined previously.
The full inference pipeline, which is used to evaluate the tumor segmentation models on the Test set, is the one illustrated in figure \ref{fig:brats-pipeline} with the trained U-Nets for tumor segmentation.
The performance of the final models on the Test set images with manually-adjusted brain extraction can be seen in table \ref{tab:auto-bet-brats-performance}.

To evaluate the impacts of a fully-automatic preprocessing in the tumor segmentation task, we evaluate the models on images preprocessed using the BE methods presented in sec. \ref{sec:brain-x}.
For this, we apply our preprocessing with each different BE method (see section \ref{sec:preprocessing}) to the Test set.
As our preprocessing returns images aligned to the BraTS images, we use the tumor annotations provided by the competition as ground truth for our evaluation.
The performance of the tumor segmentation models on images preprocessed using the different BE methods is summarized in table \ref{tab:auto-bet-brats-performance}.
A detailed view of the results can be seen in appendix \ref{apx:detailed-results}.

For comparison, we evaluate both models on images preprocessed with brain extraction performed with manual corrections.
The manually-corrected skull-stripping is achieved by extracting the brain masks of the BraTS-preprocessed images and applying these masks to the images preprocessed by us.
Therefore, by feeding this images to the tumor segmentation models, we achieve a manually-adjusted brain extraction similar to the one used by the BraTS challenge.
In table \ref{tab:auto-bet-brats-performance}, these results correspond to the "Manual" label.

\begin{table}[h]
    \center
    \caption{Median performance of the tumor segmentation models on the Test set automatically preprocessed with different BE methods. Manual represents the performance on images preprocessed with manually ajusted brain extraction. In \textbf{bold} are the best performers (excluding "Manual") under each evaluation metric.}
    \label{tab:auto-bet-brats-performance}
	\begin{tabular}{lcccc}
	\toprule
	\textbf{}           & \multicolumn{2}{c}{\textbf{2D}}                  & \multicolumn{2}{c}{\textbf{3D}}                  \\ \cline{2-5}
	\textbf{BE Method} & \textbf{Dice} $\uparrow$ & \textbf{HD95} $\downarrow$ & \textbf{Dice} $\uparrow$ & \textbf{HD95} $\downarrow$ \\ \hline
	Manual              & 0.808               & 5.242                & 0.831               & 3.977                \\
	\hline
	HD-BET              & 0.807               & 5.888                & \textbf{0.826}      & 4.616                \\
	HD-BET (fast)       & 0.806               & \textbf{5.721}       & 0.818               & \textbf{4.177}       \\
	DeepMedic           & 0.787               & 5.959                & 0.818               & 4.487                \\
	DeepMedic (multi-4) & 0.793               & 6.700                & 0.819               & 4.442                \\
	BrainMaGe           & 0.794               & 6.758                & 0.805               & 4.320                \\
	BrainMaGe (multi-4)  & 0.757              & 8.740                & 0.804               & 4.868                \\
	BET                 & 0.754               & 9.189                & 0.777               & 6.797                \\
	BET (reduce bias)   & \textbf{0.809}      & 6.250                & 0.810               & 4.608                \\
	FreeSurfer          & 0.710               & 10.492               & 0.696               & 8.828                \\
	\bottomrule
	\end{tabular}
\end{table}

\begin{figure*}[t]
    \centering
    \includegraphics[height=1.9 in]{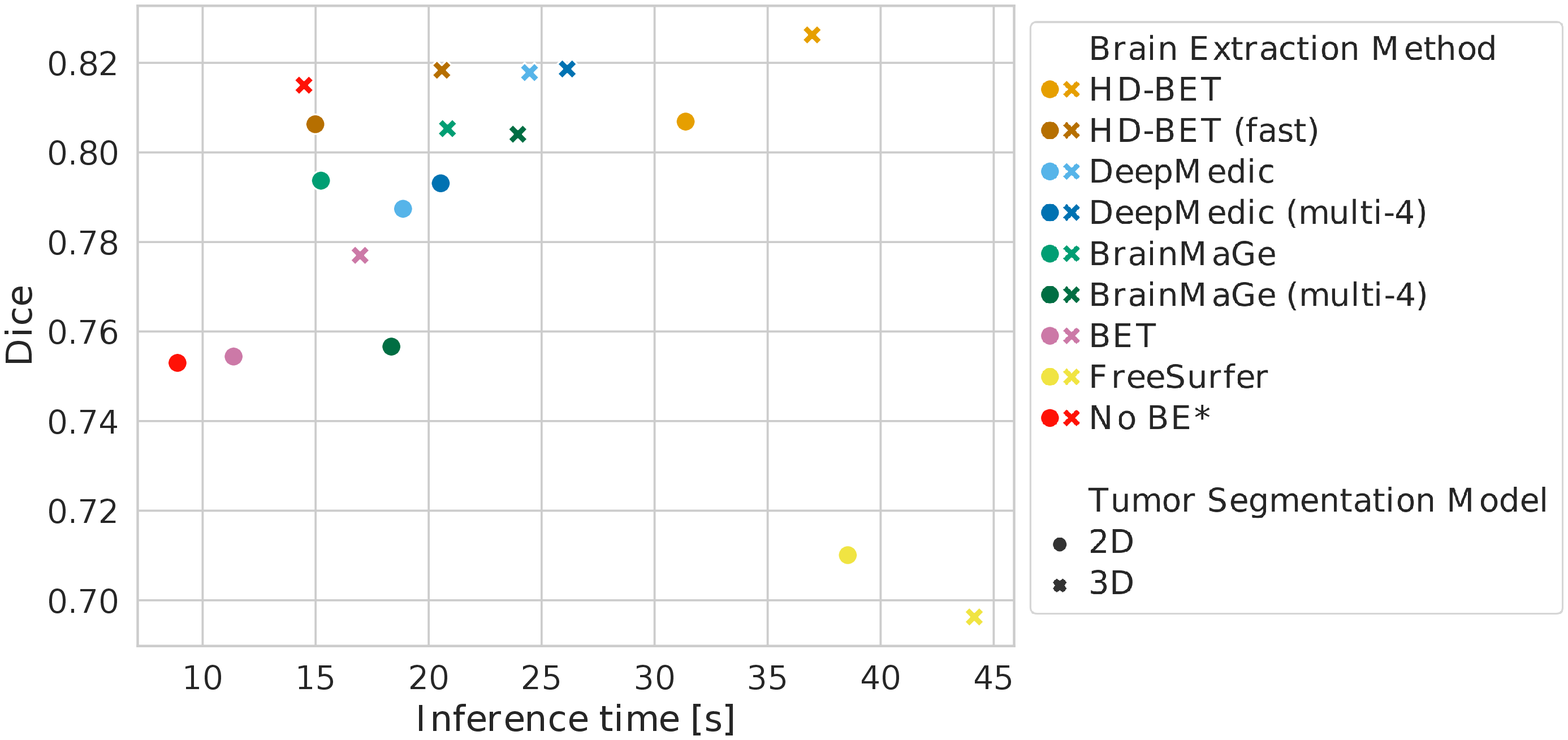}
    \includegraphics[height=1.9 in]{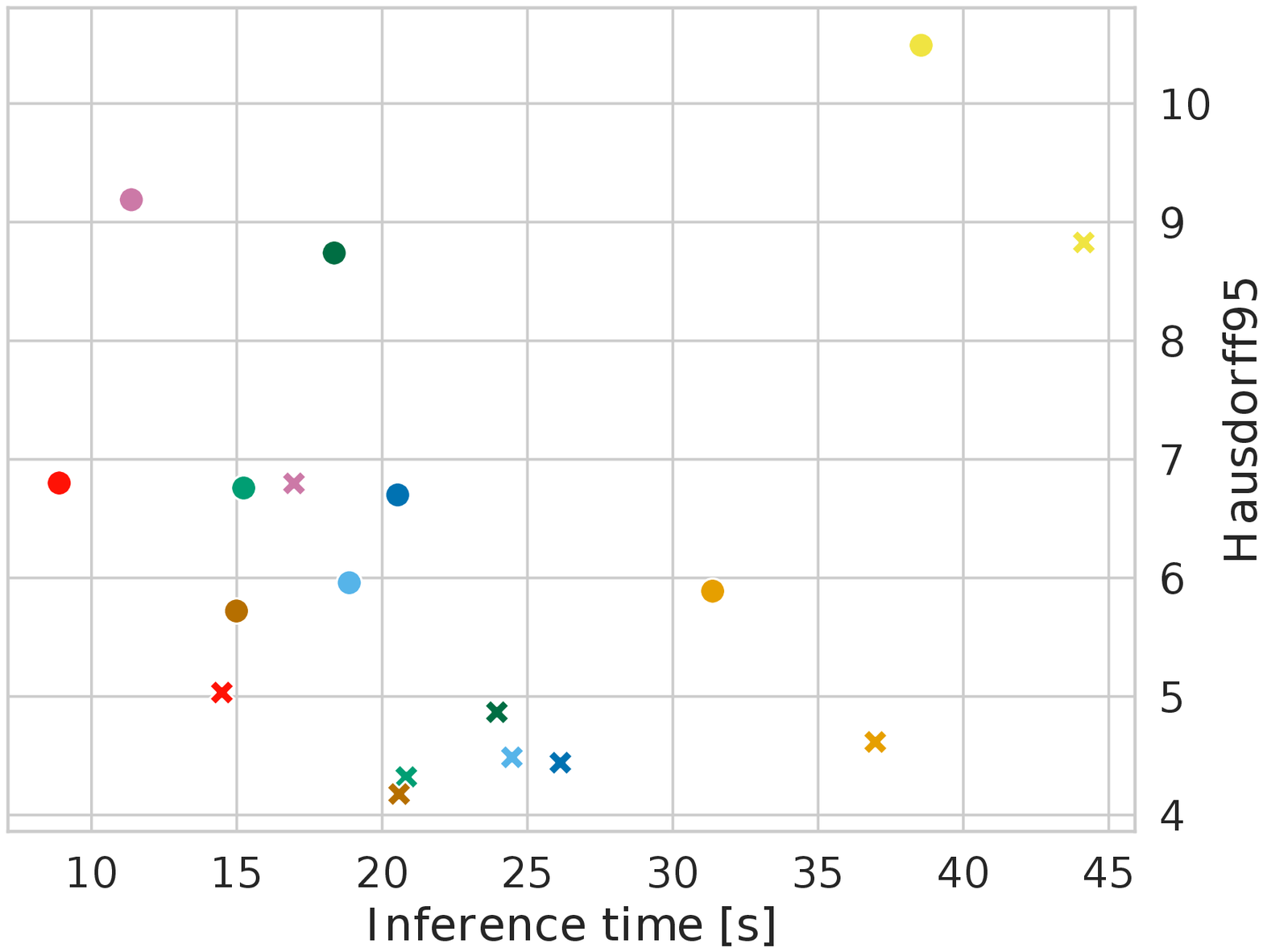}
    \caption{Tumor segmentation performance over inference time (wall time in seconds) of BE method and tumor segmentation model combinations. The color indicates which BE method was used, while the shape of the marker indicates the tumor segmentation model. *The tumor segmentation model was trained on a 55\% smaller dataset containing only non-skull-stripped images (see sec. \ref{sec:brats-no-bet}).}
    \label{fig:brats-time}
\end{figure*}

Besides the segmentation quality, we also measured the average time each BE method and each tumor segmentation model took on Test set images during inference (see sec. \ref{sec:evaluation}).
These results can be seen in table \ref{tab:inference-times}.
Furthermore, we combine the inference speed results with the tumor segmentation performance in figure \ref{fig:brats-time} to get a better view on the trade-off that exists between the two.

\begin{table}[h]
    \center
    \caption{Average inference time (wall time in seconds) of BE methods and brain tumor segmentation models on the Test set.}
    \label{tab:inference-times}
    \begin{tabular}{lcc}
	\toprule
	                      & \textbf{Average inference time} \\ \hline
	2D nnU-Net             & 8.89     \\
	3D nnU-Net             & 14.49    \\
	\bottomrule
	HD-BET                & 22.48    \\
	HD-BET (fast)         & 6.10     \\
	DeepMedic             & 9.88     \\
	DeepMedic (multi-4)   & 11.65    \\
	BrainMaGe             & 6.35     \\
	BrainMaGe (multi-4)   & 9.46     \\
	BET                   & 2.48     \\
	BET (reduce bias)     & 179.12   \\
	FreeSurfer            & 29.66    \\
	\bottomrule
    \end{tabular}
\end{table}

\subsection{Tumor segmentation without brain extraction}\label{sec:brats-no-bet}

To evaluate the effectiveness of performing BE, we train the tumor segmentation models only on non-skull-stripped images.
Therefore, we use only the subjects from the Training set for whom raw images are available (illustrated by the red highlights in fig. \ref{fig:datasets}).
Yet, we do not use any BE method, leaving the images with skull and other non-brain tissues.
For the tumor segmentation, not only do we use the same architectures, but we also apply the same procedure detailed in section \ref{sec:tumor-segmentation}.
We trained the 2D model for 100 epochs and the 3D model for 300 epochs on the images of all 167 subjects (from Training and Validation sets).

First, we train both 2D and 3D models on the images without any skull-stripping.
The test results for these two models are in table \ref{tab:no-bet} under "No BE".
Then, to benchmark these results, we train 2D and 3D models with the same hyperparameters but on the BraTS-preprocessed images (of the same subjects).
This means that we trained models following the current paradigm of applying BE to the input images, but on the same number of subjects as the proposed single-stage models, resulting in a fairer comparison.
The results of these models on the Test set, with different BE methods (similar to sec. \ref{sec:tumor-segmentation}), are on the first 9 rows of table \ref{tab:no-bet}.
A detailed view of the results can be seen in appendix \ref{apx:detailed-results}.

\begin{table}[h]
    \center
    \caption{Median Test set performance of tumor segmentation models trained only on the images from BraTS-TCGA data that are not on the Test set. \textit{No BE} are the models trained on non-skull-stripped TCGA images, while the others are models trained on skull-stripped images, similar to sec. \ref{sec:tumor-segmentation}. Manual represents the performance on images preprocessed with manually ajusted brain extraction. In \textbf{bold} are the best performers (excluding "Manual") under each evaluation metric.}
    \label{tab:no-bet}
	\begin{tabular}{lcccc}
	\toprule
	\textbf{}           & \multicolumn{2}{c}{\textbf{2D}}                  & \multicolumn{2}{c}{\textbf{3D}}                  \\ \cline{2-5}
	\textbf{} & \textbf{Dice} $\uparrow$ & \textbf{HD95} $\downarrow$ & \textbf{Dice} $\uparrow$ & \textbf{HD95} $\downarrow$ \\ \hline
	Manual              & 0.809               & 5.584                & 0.817               & 4.594                \\
	\hline
	HD-BET              & \textbf{0.808}      & 6.010                & 0.805               & \textbf{4.896}                \\
	HD-BET (fast)       & 0.807               & 5.844                & 0.804               & 5.104       \\
	DeepMedic           & 0.776               & \textbf{5.826}       & 0.807               & 5.481                \\
	DeepMedic (multi-4) & 0.773               & 6.667                & 0.803               & 5.378                \\
	BrainMaGe           & 0.772               & 7.395                & 0.783               & 8.416                \\
	BrainMaGe (multi-4) & 0.719               & 9.983                & 0.739               & 13.053                \\
	BET                 & 0.712               & 8.374                & 0.638               & 17.925                \\
	BET (reduce bias)   & 0.795               & 6.449                & 0.803               & 5.652                \\
	FreeSurfer          & 0.683               & 15.244               & 0.628               & 19.229                \\
	\hline
	No BE & 0.753               & 6.800                & \textbf{0.815}      & 5.033                \\
	\bottomrule
	\end{tabular}
\end{table}

As the same architectures were used, the inference times of the tumor segmentation models trained without skull-stripping are the same as those trained with skull-stripping (in sec. \ref{sec:tumor-segmentation}).
Still, as no BE is necessary, the whole pipeline runs significantly faster.
The performance of the models trained without skull-stripping are competitive even to the models from the experiments from sec. \ref{sec:tumor-segmentation}, as figure \ref{fig:brats-time} shows, which had a training set more than twice as big (369 subjects against 167).

\section{Discussion}

Even though BE is a common preprocessing operation for brain tumor segmentation, its impact on the tumor segmentation quality has never been measured before.
Our results show that a poorly selected BE method can compromise up to 13 p.p. (HD-BET in comparison to FreeSurfer) of tumor segmentation Dice score.
Furthermore, HD-BET was the method closest to the manually-adjusted, gold-standard BE, having effectively no negative impact on the tumor segmentation.

Our preprocessing pipeline applied to the raw images from BraTS-TCGA can effectively align them to the tumor annotations provided by BraTS, as the results in sec. \ref{sec:preprocessing-experiments} show.
This enabled the first evaluation of a fully-automatic brain tumor segmentation pipeline (including preprocessing), which would suit the needs of a clinical setup.

Furthermore, our experiments with BE agree with the literature in regard to the superiority of deep-learning-based methods over traditional algorithms \cite{Isensee2019AutomatedNetworks,Thakur2020BrainTraining}.
The only surprise was BET with the "reduce bias" option, which achieved a high performance (but with a very high inference time).
The performance of 3D models for tumor segmentation in comparison to 2D models also follows the trends seen in BraTS results from recent years.

Still, as deep learning models for both tasks (tumor segmentation and BE) need powerful hardware to perform inference in reasonable time, not always the use of state-of-the-art models for both tasks is possible.
In a scenario with a limited computing budget, it would seem intuitive to opt for the best tumor segmentation model possible and then pick the BE method that still fits in the budget.
Yet, as figure \ref{fig:brats-time} illustrates, if low inference time (under 20 seconds) is a requirement, the choice to use the 2D nnU-Net with HD-BET in its fast configuration yields much better results than 3D nnU-Net with BET.
This shows that the best tumor segmentation model is not always the best choice for low budget scenarios, as its performance can be deeply hindered by the BE method employed.

Furthermore, our experiments with the proposed single-stage models, trained on images without skull-stripping, indicate that performing BE might not be necessary at all.
Unfortunately, a large and multi-institutional dataset of raw images with tumor annotations is not currently available, making a proper comparison impossible.
Nevertheless, the 3D model trained on non-skull-stripped images achieved results on par with the state-of-the-art tumor segmentation approach, while using a smaller training set.
However, the single-stage approach performs inference much faster.
Thus, with more data and proper hardware, a tumor segmentation pipeline without a BE step could be a suitable option, as it can provide high segmentation quality at shorter inference times.

Nevertheless, no automatic approach was able to overcome the results achieved by manually correcting the brain masks.
Namely, our results agree with those reported in the BraTS competition \cite{Bakas2018IdentifyingChallenge} in that 3D tumor segmentation models trained on images preprocessed with manual interventions achieve result in tumor segmentations closer to the ones performed by experts.
We note, however, that the performance of fully-automatic approaches is very close to that of those with manual interventions.
Therefore, we speculate that better results may be achieved if the expert's time is invested in improving the tumor segmentation resulting from a fully-automatic pipeline rather than fixing the brain mask before running the tumor segmentation model.

\begin{figure*}[T]
    \centering
    \begin{subfigure}[b]{\textwidth}
        \centering
        \includegraphics[width=0.45\textwidth]{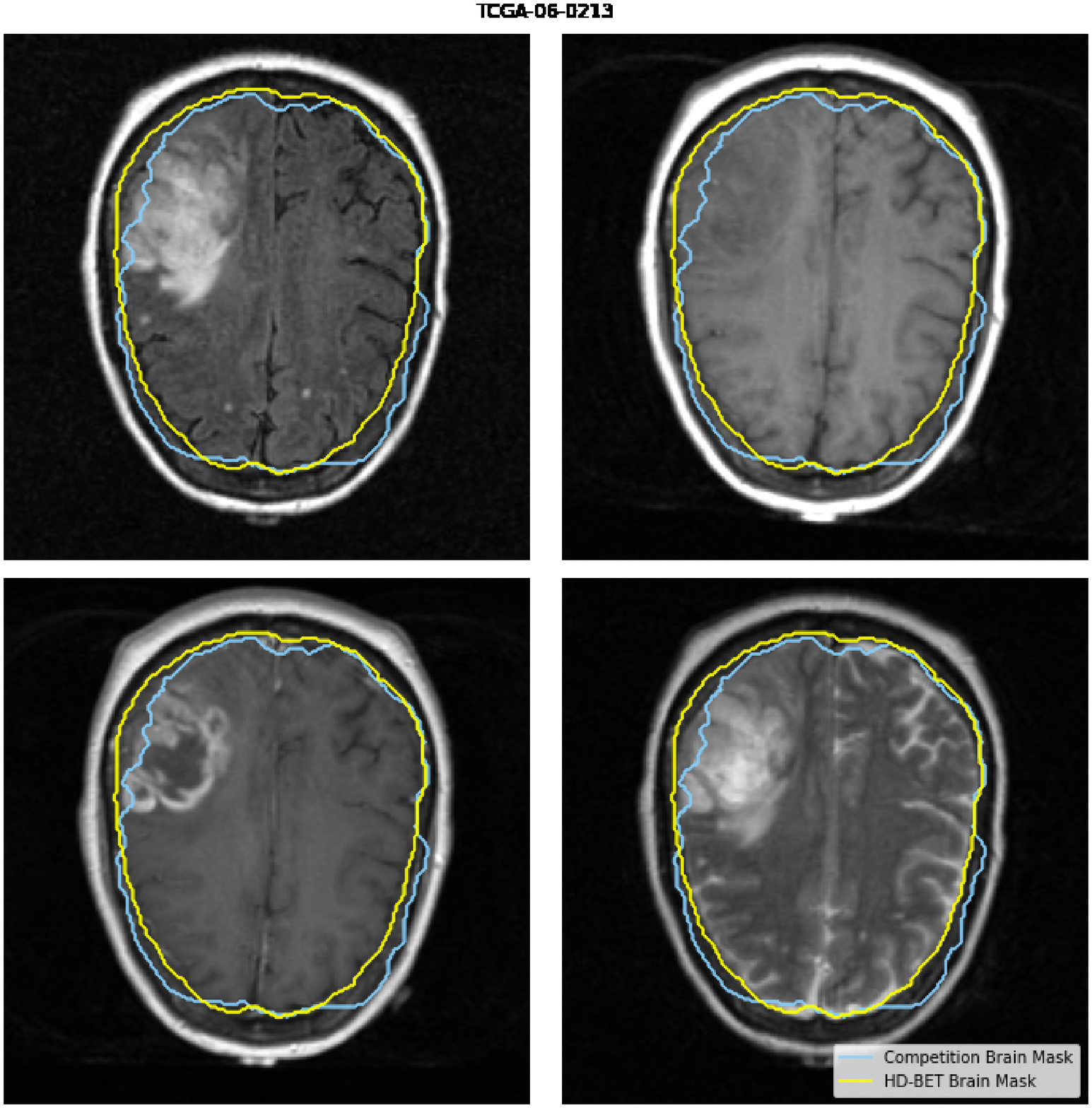}
        \includegraphics[width=0.45\textwidth]{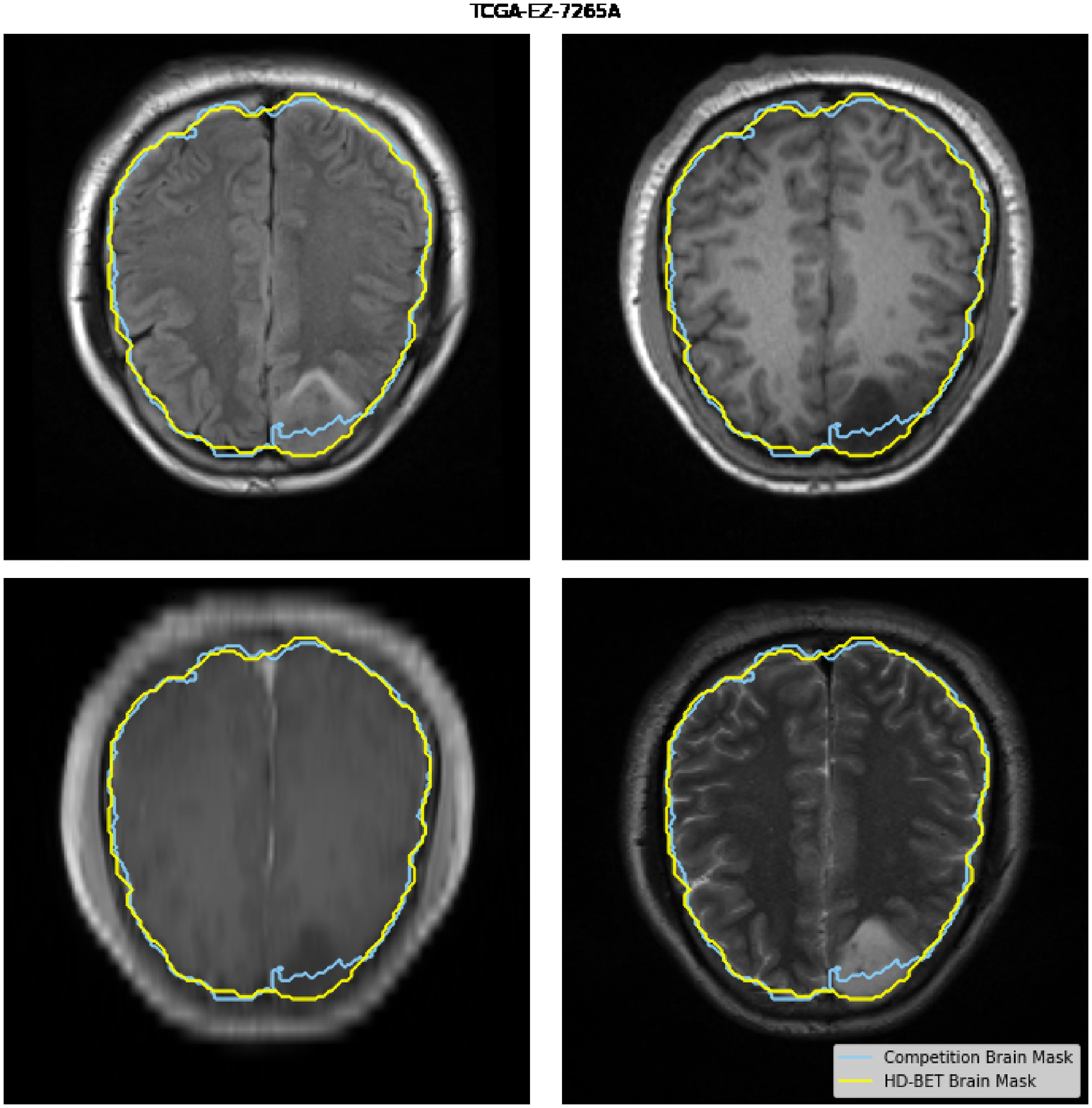}
	\caption{}\label{fig:brats-examples-a}
    \end{subfigure}
    \hfill
    \begin{subfigure}[b]{\textwidth}
        \centering
        \includegraphics[width=0.45\textwidth]{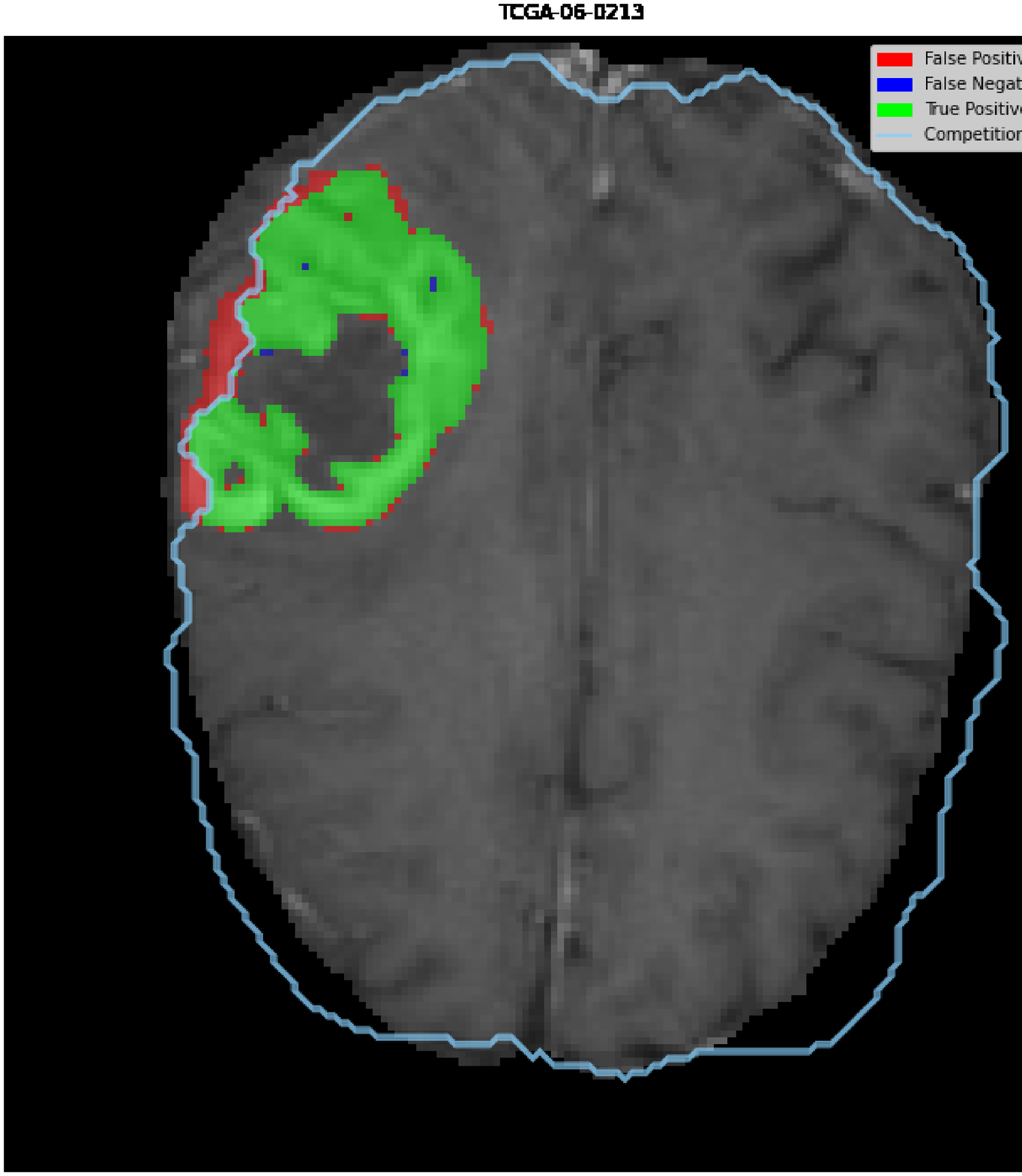}
        \includegraphics[width=0.45\textwidth]{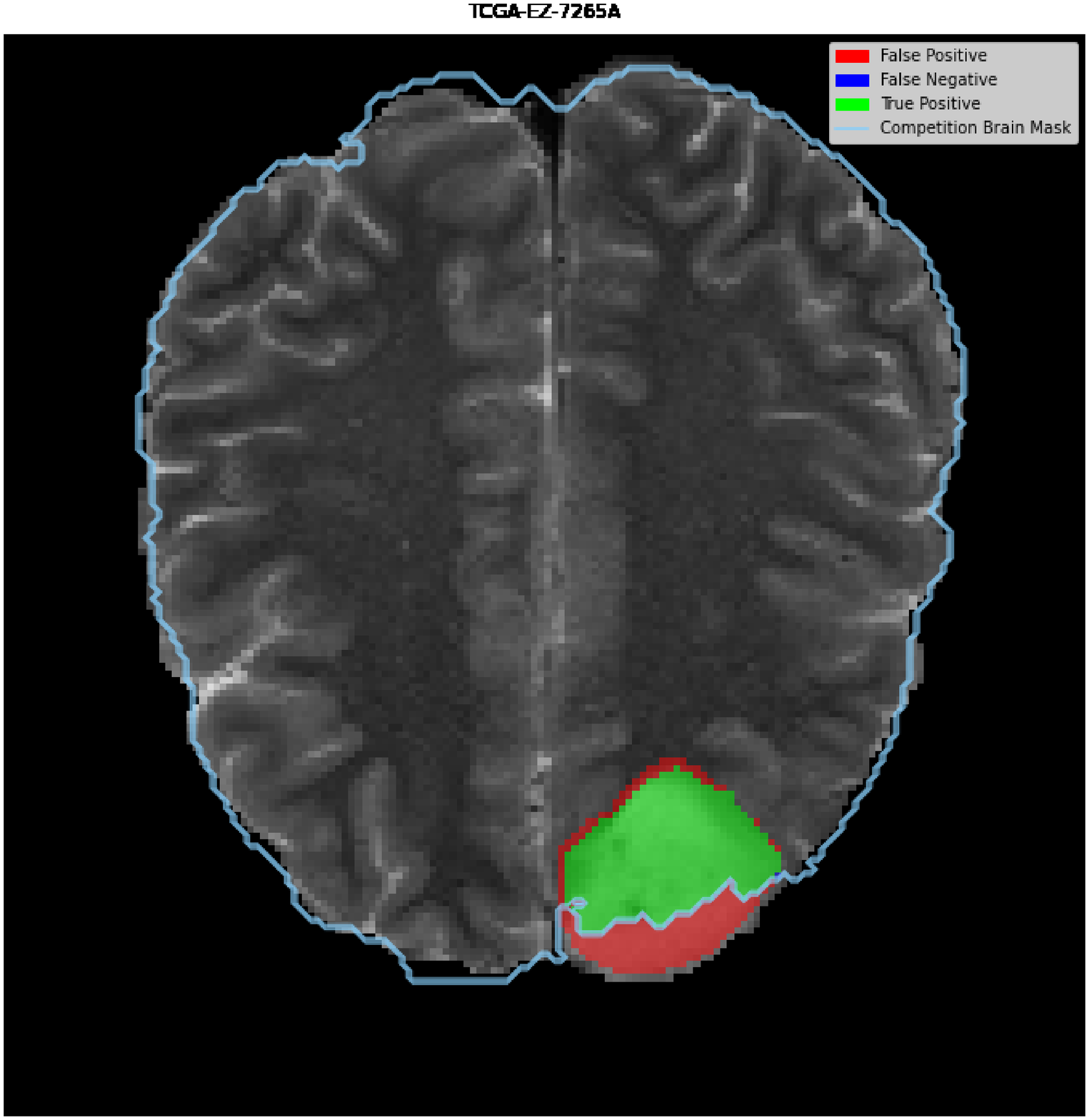}
	\caption{}\label{fig:brats-examples-b}
    \end{subfigure}
    \caption{Differences between the brain mask in the BraTS-preprocessed images and the one generated by HD-BET on two subjects. In \textit{(a)}, the contour of the masks are overlapped on the four modalities of each subject. In \textit{(b)}, the evaluation of the segmentation generated by the 3D nnU-Net on both images (BraTS-preprocessed and preprocessed using HD-BET) against the ground truth is illustrated for ET region on the left and TC region on the right.}
    \label{fig:brats-examples}
\end{figure*}

A closer look at the images preprocessed using HD-BET (the BE method with the lowest overall impact on the tumor segmentation) shows that they are very close to the BraTS-preprocessed images.
Still, as visible in figure \ref{fig:brats-examples-a}, some images differ significantly in the tumor region, which is the most sensitive for the tumor segmentation.
Some images preprocessed by the competition do not contain all the tumorous information of the original image.
Since the ground truth was generated on the preprocessed images rather than on the raw images, such deviations hinder the quality of the evaluation.
Note the examples in figure \ref{fig:brats-examples-b}, where the tumor segmentation performed in the image preprocessed using HD-BET has plenty of correctly segmented tumorouos voxels labeled as false positives.

\section{Conclusions}

In this study, we have presented, for the first time, an evaluation of BE methods as preprocessing steps for the tumor segmentation task.
Our results show that a poorly selected BE method can reduce up to 15.7\% of the tumor segmentation performance\footnote{Measured through the Dice score for the 3D tumor segmentation model, as presented in table \ref{tab:auto-bet-brats-performance}}.

On the other hand, our proposed single-stage approach of training brain tumor segmentation deep learning models on non-skull-stripped images has shown to be very promising, achieving excellent results even though many fewer images were available for training when compared to state-of-the-art models.
Even though BE is still necessary for many brain-related studies, our results suggest that the current paradigm of training tumor segmentation models on skull-stripped images \cite{Bakas2018IdentifyingChallenge} might not be optimal for developing models suitable for clinical practice.

Our experiments with the preprocessing of BraTS-TCGA images as well as our error analysis highlighted certain improvements that would be needed in brain tumor datasets for the benchmarking of a complete brain tumor segmentation pipeline.
A large, multi-institutional brain imaging dataset with gold-standard tumor annotations aligned to the raw images, if available, could help improve the reliability of the evaluation of BE methods and consolidate single-stage tumor segmentation models.
Until then, manually-adjusted BE still provides the most reliable results, even though automatic BE methods achieve comparable results.

\section*{Declaration of competing interest}

The authors declare no conflicts of interest.

\printcredits

\bibliographystyle{model1-num-names}

\bibliography{references}

 


\appendix

\section{Performance of tumor segmentation models with different BE methods}\label{apx:detailed-results}

\begin{figure*}[H]
    \centering
    \begin{subfigure}[b]{.9\textwidth}
        \centering
        \includegraphics[width=\textwidth]{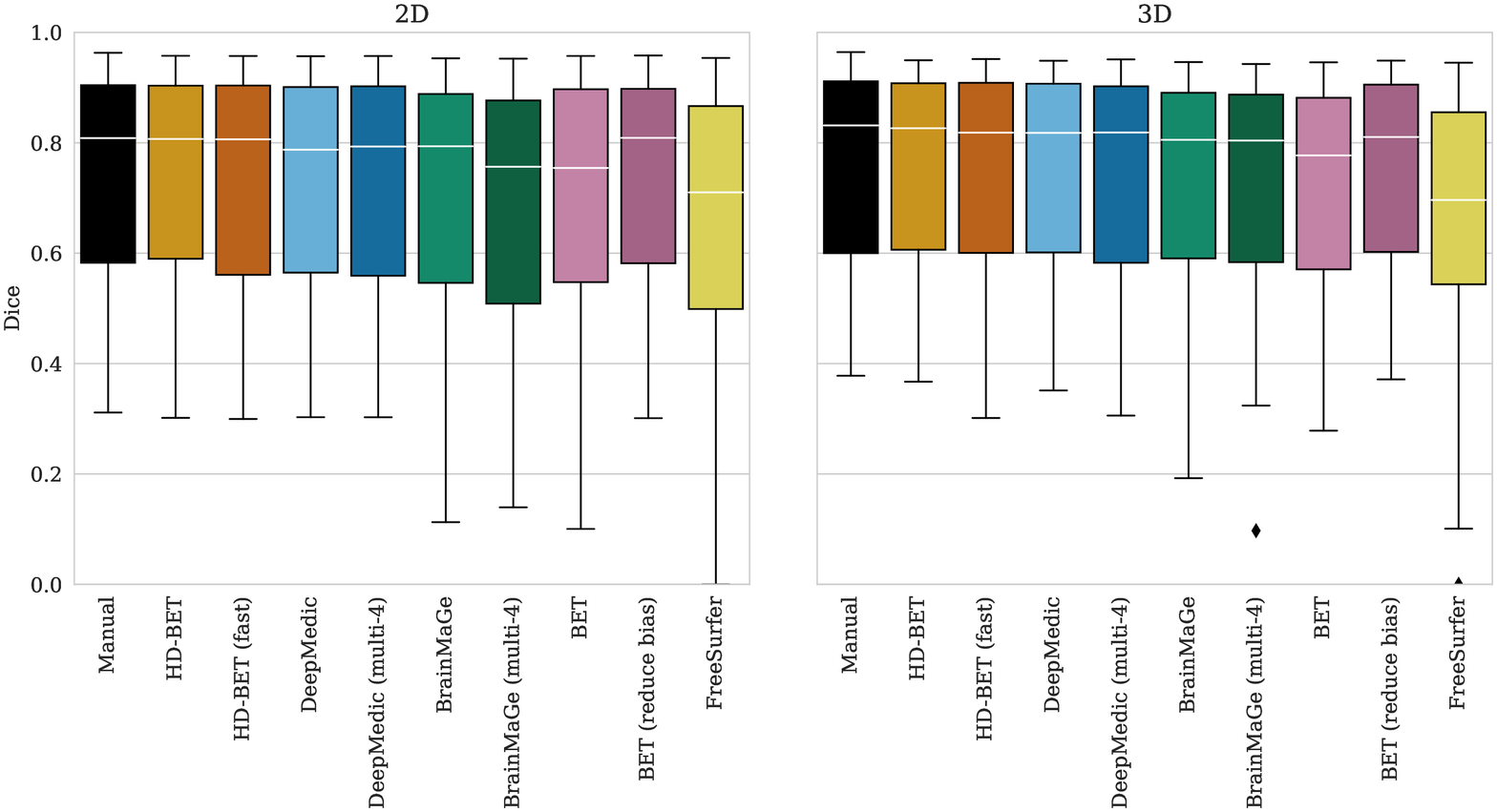}
    	\caption{}\label{fig:be-boxplot-dice}
    \end{subfigure}
    \begin{subfigure}[b]{.9\textwidth}
        \centering
        \includegraphics[width=\textwidth]{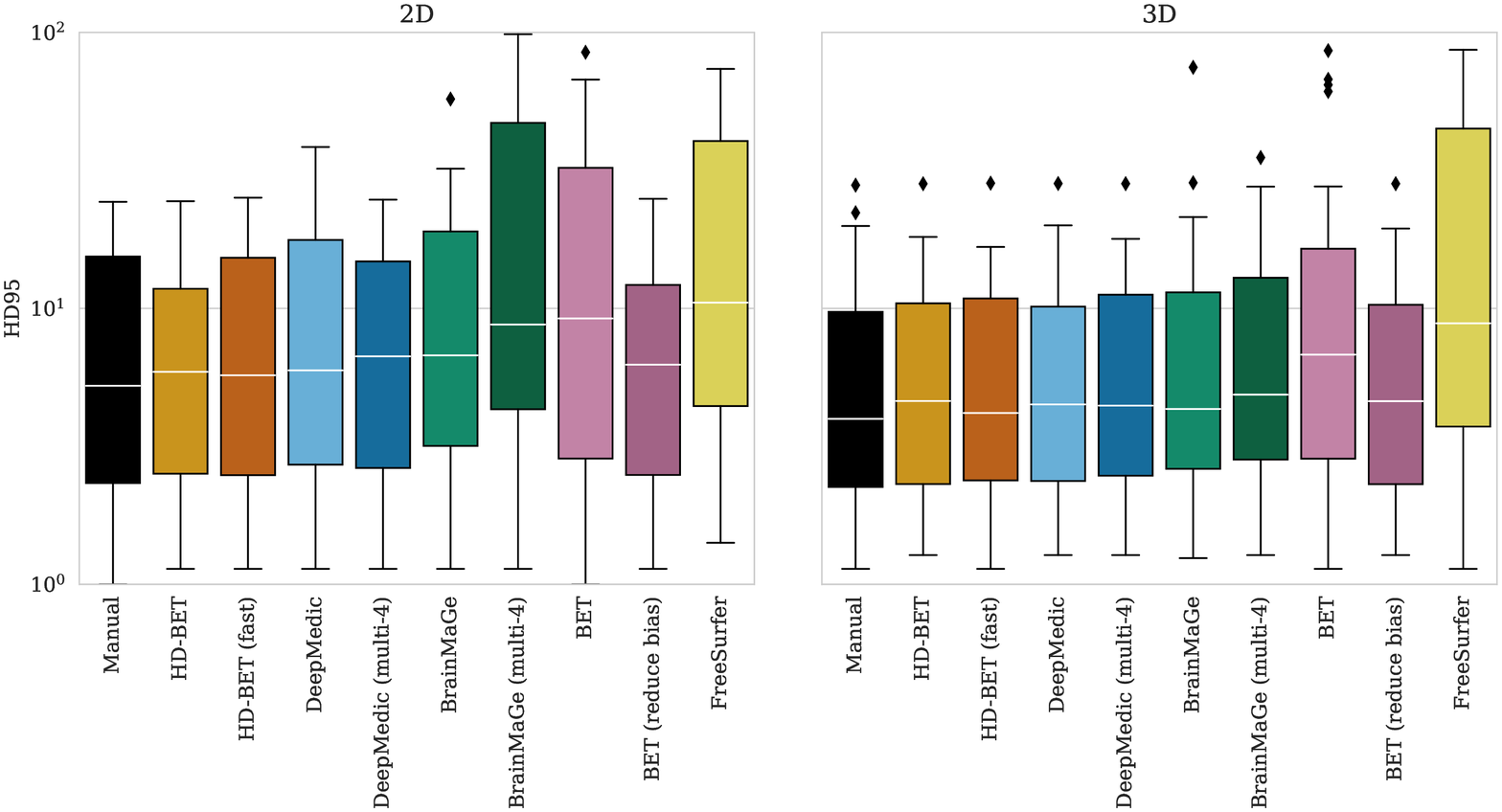}
    	\caption{}\label{fig:be-boxplot-hd95}
    \end{subfigure}
    \caption{Test set performance of tumor segmentation models with different BE methods, as measured by the Dice score ($a$) and the Hausdorf95 ($b$). \textit{Manual} represents the performance on images preprocessed with manually adjusted brain extraction.}
    \label{fig:be-boxplot}
\end{figure*}

\begin{figure*}[H]
    \centering
    \begin{subfigure}[b]{.9\textwidth}
        \centering
        \includegraphics[width=\textwidth]{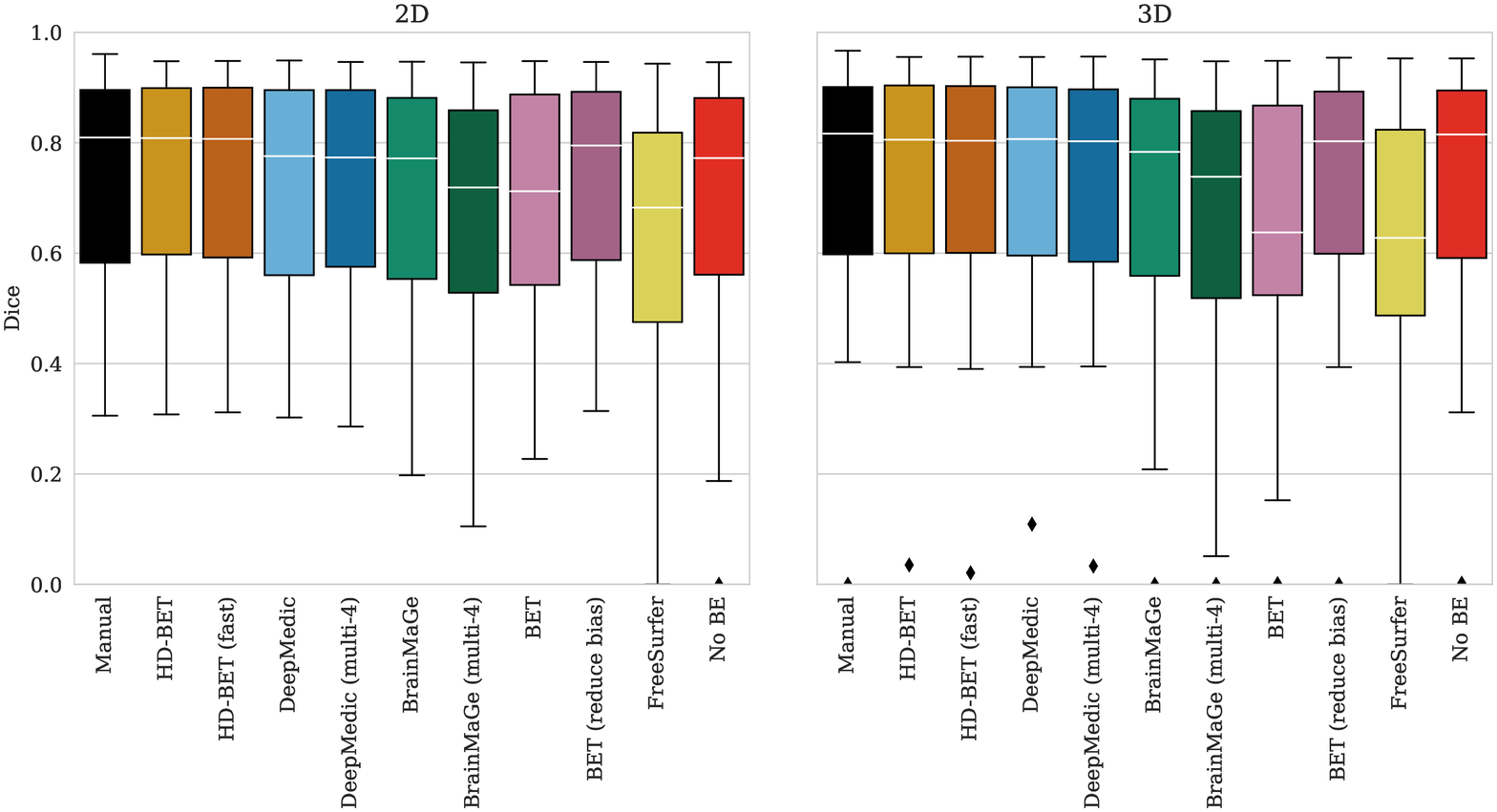}
    	\caption{}\label{fig:nobe-boxplot-dice}
    \end{subfigure}
    \begin{subfigure}[b]{.9\textwidth}
        \centering
        \includegraphics[width=\textwidth]{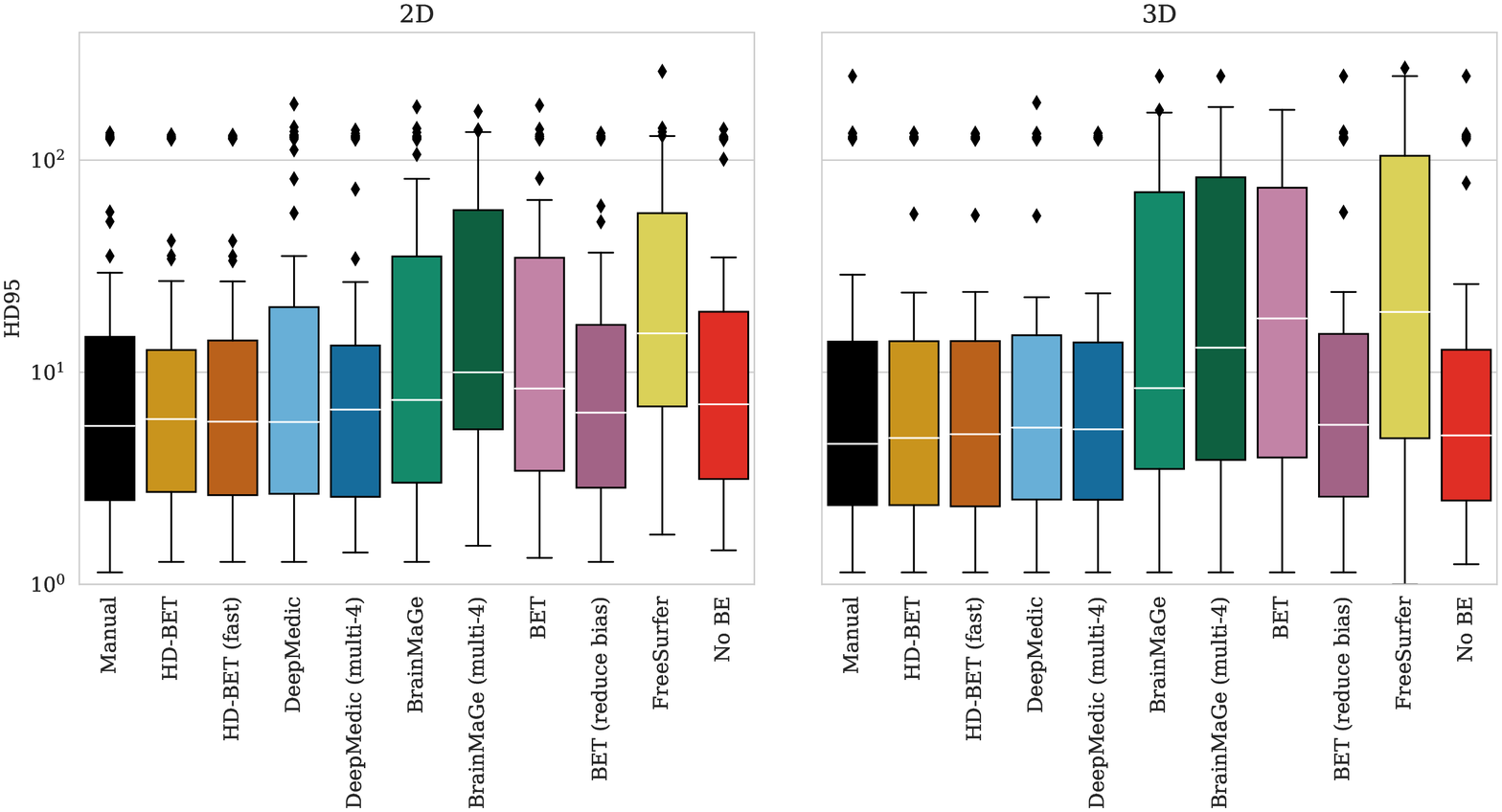}
    	\caption{}\label{fig:nobe-boxplot-hd95}
    \end{subfigure}
    \caption{Test set performance of tumor segmentation models trained only on images from BraTS-TCGA data, as measured by the Dice score ($a$) and the Hausdorf95 ($b$). \textit{Manual} represents the performance on images preprocessed with manually adjusted brain extraction. \textit{No BE} are the models trained on non-skull-stripped images.}
    \label{fig:nobe-boxplot}
\end{figure*}

A detailed description of the Test set results summarized in tables \ref{tab:auto-bet-brats-performance} and \ref{tab:no-bet} is presented in here.
More precisely, the performance of the tumor segmentation models with different BE methods, as described in sec. \ref{sec:tumor-segmentation}, is illustrated through the box plots of fig. \ref{fig:be-boxplot}, while the performance of the tumor segmentation models trained only on subjects for which the raw images are available, as described in sec. \ref{sec:brats-no-bet}, is illustrated through the box plots of fig. \ref{fig:nobe-boxplot}.

\end{document}